\documentclass{elsart}
\usepackage{epsfig}
\usepackage{amssymb}
\makeatletter
\renewcommand\theequation{\thesection.\arabic{equation}}
\@addtoreset{equation}{section}
\makeatother
\begin{document}
\begin{frontmatter}
\title{Light-Ion-Induced Multifragmentation: \\
 The ISiS Project}
\author[Indiana]{V.E. Viola},
\author[LANL]{K. Kwiatkowski$^{a,}$},
\author[Laval]{L. Beaulieu$^{a,}$},
\author[LANL]{D.S. Bracken$^{a,}$},
\author[Maryland]{H. Breuer},
\author[Jagellonian]{J. Brzychczyk$^{a,}$},
\author[Indiana]{R.T. de Souza},
\author[Washington]{D.S. Ginger$^{a,}$},
\author[Indiana]{W-c. Hsi},
\author[SimonFraser]{R.G. Korteling},
\author[Caen]{T.Lefort$^{a,}$},
\author[MSU]{W.G. Lynch},
\author[LANL]{K.B. Morley$^{a,}$},
\author[deceased]{R. Legrain$^{\ell,}$}, 
\author[Warsaw]{L. Pienkowski},
\author[CEA]{E.C. Pollacco},
\author[Microsoft]{E. Renshaw$^{a,}$},
\author[Bangkok]{A. Ruangma$^{m,}$}
\author[MSU]{M.B. Tsang},
\author[CEA]{C. Volant}, 
\author[Systems]{G. Wang$^{a,}$},
\author[TexasAM]{S.J. Yennello} 
\address[Indiana]{Indiana University,Bloomington, IN 47405} 
\address[LANL]{present address: Los Alamos
National Laboratory, Los Alamos, NM 87545}
\address[Laval]{present address: Laval University, Quebec City, Canada G1K 7P4}
\address[Maryland]{University of Maryland, College Park, MD 20740}
\address[Jagellonian]{Jagellonian University, Krakow, Poland}
\address[Washington]{present address: University of Washington,
Seattle, WA} 
\address[SimonFraser]{Simon Fraser University, Burnaby
BC, Canada V5A IS6} 
\address[Caen]{present address: University of Caen, Caen,
France}
\address[MSU]{National Superconducting Laboratory, Michigan State University, East Lansing, MI 48824}
\address[CEA]{CEA Saclay, Saclay, France}
\address[Warsaw]{Warsaw University, Warsaw, Poland}
\address[Microsoft]{present address: Microsoft Corporation,Redmond, WA 98052}
\address[TexasAM]{Texas A\& M University, College Station, TX 77843}
\address[Bangkok]{present address: Siriraj Hospital, Bangkok, 10700 Thailand}
\address[Systems]{present address: Epsilon Corp., Irving, TX}
\address[deceased]{deceased}
\begin{abstract}
An extensive study of GeV light-ion-induced multifragmentation and its
possible interpretation in terms of a nuclear liquid-gas phase
transition has been performed with the Indiana Silicon Sphere (ISiS)
4$\pi$ detector array. Measurements were performed with 5-15 GeV/c p,
$\overline{p}$, and $\pi^-$ beams incident on $^{197}$Au and 2-5 GeV $^3$He
incident on $^{nat}$Ag and $^{197}$Au targets.  Both the reaction
dynamics and the subsequent decay of the heavy residues have been
explored.  The data provide evidence for a dramatic change in the
reaction observables near an excitation energy of E*/A = 4-5 MeV per
residue nucleon.  In this region, fragment multiplicities and energy
spectra indicate emission from an expanded/dilute source on a very
short time scale (20-50 fm/c).  These properties, along with caloric
curve and scaling-law behavior, yield a pattern that is consistent
with a nuclear liquid-gas phase transition. 
\end{abstract}

\end{frontmatter}

\noindent Contents

1.  Introduction

2.  The ISiS Experimental Program

3.  Reaction Dynamics

\hspace{5mm}3.1  Excitation Energy Deposition

\hspace{5mm}3.2  BUU Simulations

\hspace{5mm}3.3  Sideways Peaking

4.  Statistical Decay Multifragmentation

\hspace{5mm}4.1	 Calorimetry

\hspace{5mm}4.2  Thermal Observables

\hspace{9mm}4.2.1 Fragment Spectra

\hspace{9mm}4.2.2 Multiplicities

\hspace{9mm}4.2.3 Charge Distributions

\hspace{9mm}4.3.4 Cross Sections

\hspace{9mm}4.2.5 Source Charge

\hspace{5mm}4.3	 Breakup Density and Expansion

\hspace{5mm}4.4  Breakup Time Scale

5.	Thermodynamics

\hspace{5mm}5.1 The Caloric Curve:  Isotope-ratio Temperatures

\hspace{5mm}5.2 The Caloric Curve:  Density-dependent Fermi Gas Temperatures

\hspace{5mm}5.3 Heat Capacity

6.	The Liquid-Gas Phase Transition:  Scaling Law Behavior

7.	Summary and Conclusions

8.      Acknowledgements


\section{Introduction}

One of the most important signals of the formation of hot nuclear
matter is the emission of nuclear clusters, or intermediate-mass
fragments (IMF: 3 $\leq$ Z $\lesssim$ 20).  From studies of the IMF
yields in energetic nuclear reactions one hopes to gain greater
insight into the thermodynamics of highly-excited nuclei and the
nuclear equation-of-state at low densities.  An important aspect of
such studies is the identification of a possible nuclear liquid-gas
phase transition \cite{Fi82,Be83,Bon85}.

IMF emission was first observed in the 1950's, when beams of protons
and alpha particles with energies in the GeV range became available.
Emulsion and radiochemical measurements of the reaction products
showed that the probability for cluster emission increased strongly
with beam energy, suggesting their association with the decay of
highly excited nuclei.  The emulsion measurements also provided
evidence for multiple fragment emission, or multifragmentation.  These
data led to the concept that the reaction mechanism could be
schematically viewed in terms of a two-step intranuclear
cascade/statistical emission model, with IMFs emitted primarily 
in the evaporation stage.
One complication with this model was the observation that lighter
clusters exhibited forward-peaked angular distributions,
suggesting that a prompt non-statistical mechanism must also be
present.  Another perceptive proposal was that pion production and
reabsorbtion could be a major mechanism for excitation-energy
deposition in the heavy residual nucleus\cite{Fr54,Wo56}.  These
studies, carried out at Lawrence Berkeley Laboratory, Brookhaven
National Laboratory and in Russia, laid the groundwork for future
studies and are reviewed in \cite{Pe60,Cr63,Hu67,Ly87}.

The development of silicon semiconductor detectors in the 1960's
 made it possible to perform measurements of inclusive IMF yields and kinetic
energy spectra. Bombardments of heavy targets
with 5 GeV protons provided a more systematic understanding of the
earlier work, and showed clearly the existence of two mechanisms for
IMF production -- one equilibrium-like and the other a fast
nonequilibrium process \cite{Po71,Ko73}.  An important aspect 
of the spectra was a
downward shift in the apparent Coulomb barrier relative to lower
energy reactions.  It was suggested that this shift was due to a
modified density distribution, perhaps due to an expanded source.
Later, similar inclusive measurements with protons up to 350
GeV \cite{Hi84} confirmed radiochemical studies that indicated
the IMF emission probability reaches a maximum near 10 GeV and remains
constant thereafter \cite{Po89}.  These later measurements also tracked
the evolution of the Coulomb peak displacement with beam energy, and found
its onset to be near 4 GeV. This result was interpreted as a possible sign of
critical behavior \cite{Fi82,Pa84}, and stimulated widespread interest
in this possibility.

To search for more conclusive evidence of a phase transition and
critical behavior, it was necessary to develop large-solid angle
detector arrays to provide fragment kinetic-energy spectra,
multiplicity information, event topology and calorimetry.  Among the
most important features of such a device are: (1) nuclide (Z and A)
identification of all products, including neutrons, (2) spatial
characterization with good granularity; (3) low detection thresholds;
(4) good energy resolution over a large dynamic range and (5)
efficient, reliable detector calibration techniques. Another important
factor in detector design is the geometry imposed by reaction
kinematics; e.g. IMF fragment distributions are nearly isotropic in
light-ion-induced reactions, with a small component in the beam
direction, whereas they are strongly forward-focused in heavy-ion
reactions.

With improvements in detector and data-acquisition techniques, it
became possible to construct complex detector arrays for performing
exclusive measurements that met most, but not all, of the above
conditions for complex fragments. The first such array was the LBL
Plastic Ball/Wall \cite{Ba82}, which demonstrated the existence of
events with high multiplicities of IMFs and light-charged particles
(LCP: Z = 1,2) \cite{Wa83}.  Later measurements at LNS Saclay
confirmed the high IMF multiplicities and yield dependence on
bombarding energy \cite{Ye93}. These measurements also showed that the
IMF spectral peak energies decreased as a function of increasing
multiplicity, consistent with the expansion scenario.  Subsequently,
several 4$\pi$ detector arrays were constructed for the study of
light-ion induced multifragmentation: the EOS TPC \cite{EOS}, the
Berlin Silicon/Neutron Ball \cite{BNB}, the Dubna FASA array
\cite{Av93} and the Indiana Silicon Sphere (ISiS) \cite{Kw95a}.  At
the same time heavy-ion accelerator technology had advanced to the
stage where it became possible to study multifragmentation in A + A
reactions, which lead to the construction of several additional
detectors \cite{Ce91,De90,Pou95,NIMROD,Ly89,Gulminelli}. At the outset
it should be stressed that there is general concordance among the
results of all of these projects with those from ISiS, as can be found
in \cite{Gulminelli}. In the following sections, we survey the salient
results of the ISiS program with GeV light-ion projectiles, and
address the question: is there evidence for a nuclear liquid-gas phase
transition?

\section{The ISiS Experimental Program}

The Indiana Silicon Sphere project was initiated in the late
1980's with a specific focus on GeV light-ion-induced reactions on
heavy nuclei. Light ions bring a unique perspective to
multifragmentation studies in that they emphasize the thermal
properties of the disintegrating residue, with minimal distortions of
the spectra due to rotational and compressional effects.
Experimentally, GeV light ion beams form only one emitting source,
which decays in a reference frame that is very close to that of the
source frame (source velocities $\sim$ 0.01c).  In addition, the
energy-deposition mechanism produces a continuous distribution
of excitation energies, permitting broad coverage of the excitation
energy distribution in a single reaction.

\begin{figure}
\vspace{25mm}
\centerline{\psfig{file=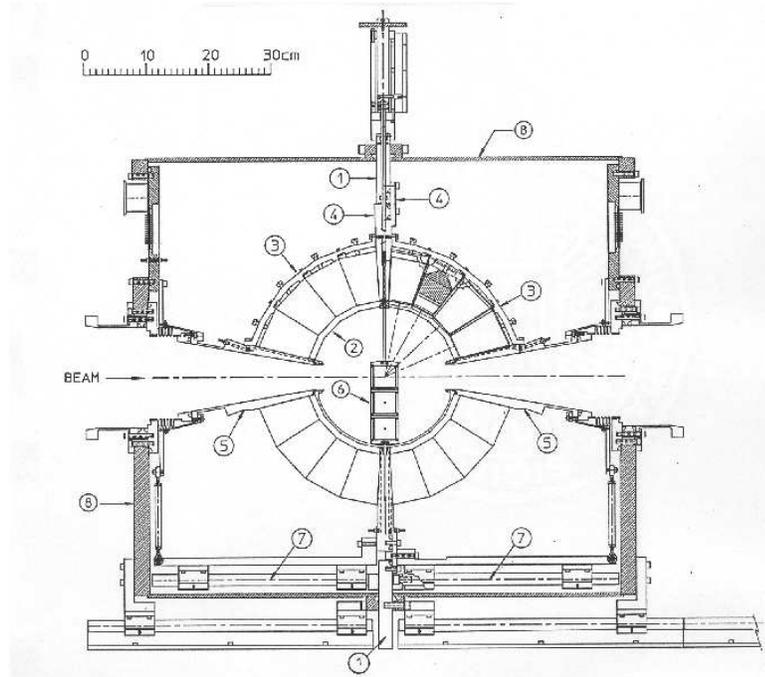,width=4.0in}}
\caption{Assembly drawing of the ISiS detector array.  Components are
as follows: (1) center support ring; (2) gas-vacuum separation window;
(3) arc support bars; (4) partition disks; (5) beamline support cones;
(6) target ladder assembly; (7) steel rails for opening housing cans
and (8) vacuum chamber.}
\label{Fig1}
\end{figure}

\begin{figure}
\vspace{20mm}
\centerline{\psfig{file=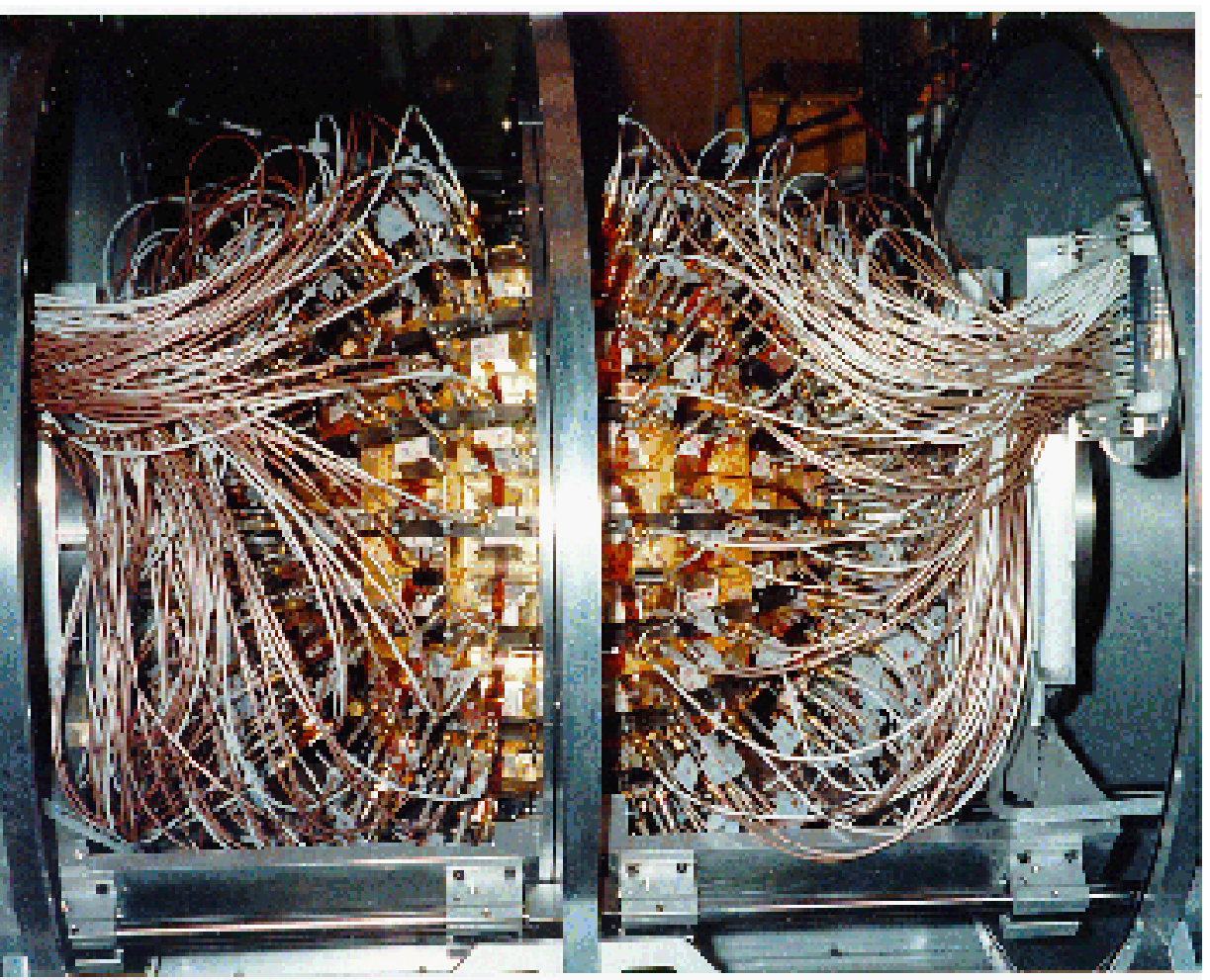,width=4.0in}}
\caption{Photograph of the ISiS detector loaded with detector telescopes.}
\label{Fig2}
\end{figure}

Based upon previous inclusive and semi-exclusive studies \cite{Ye93},
the ISiS 4$\pi$ array was designed in a spherical geometry, with very
low detector thresholds and a large dynamic range for LCPs [LCP = H
and He] and IMFs \cite{Kw95a}.  Particle identification was based on
triple telescopes composed of an ion-implanted silicon detector, preceded
by a gas-ion chamber for low energy fragments and followed by a CsI
(T$\ell$) crystal for the most energetic particles, primarily LCPs.
An overall schematic diagram of ISiS is shown in Fig.\ \ref{Fig1},
accompanied by a photo in Fig. \ \ref{Fig2}.  The array consists of
162 particle-identification telescopes, 90 covering the angles
14$^{\circ}$-86.4$^{\circ}$ and 72 spanning
93.6$^{\circ}$-166$^{\circ}$. The telescopes are arranged in eight
rings, each of which is composed of 18 truncated ion-chamber housings.
In the forward-most ring, the Si and CsI detectors were divided into
two segments to increase granularity.  Detector acceptance was 74\% of
4$\pi$.

\begin{figure}
\vspace{20mm}
\centerline{\psfig{file=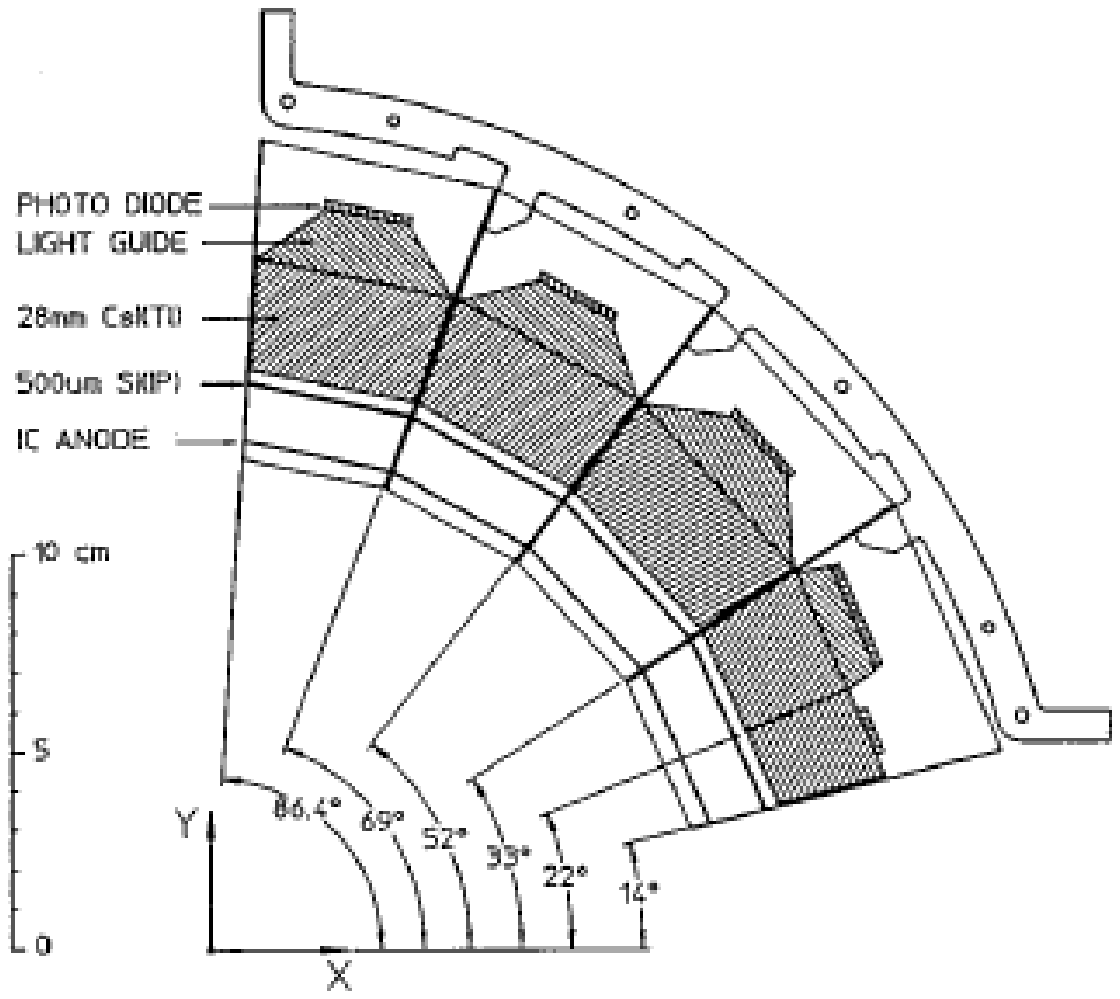,width=4.0in}}
\caption{Drawing of an ISiS arc bar segment for the forward hemisphere with
the angular coverage of each telescope labeled.  Each segment is part of
an 18-member ring; the forward-most element is divided into two
segments.  Forward-angle rings are identified as follows: 14-22$^\circ$;
22-33$^\circ$; 33-52$^\circ$; 52-69$^\circ$;
69-86.4$^\circ$.  Backward-angle rings are:  93.6-111$^\circ$;
111-128$^\circ$; 128-147$^\circ$; and 147-166$^\circ$.}
\label{Fig3}
\end{figure}

A schematic drawing of the detector telescopes is shown in
Fig. \ref{Fig3}.  The detectors are mounted in gold-plated,
high-conductivity copper cans.  The first element in each telescope
($\Delta$E) is an axial-field gas-ionization chamber (GIC), operated
at $\sim$ 200 V and a pressure of 15-20 Torr of C$_3$F$_8$ gas.  These
conditions permitted identification of fragments with energies as low
as $\sim$ 0.8 MeV/nucleon.  All detectors operate in a common
gas volume in each hemisphere, with vacuum isolation provided by a
$\sim$ 120-150 $\mu$g/cm$^2$ stretched polypropylene window covered with a thin
graphite coating.  The ion chambers are followed by an ion-implanted
silicon detector of thickness 500 $\mu$m, which is sufficient to stop
E/A $\sim$ 8 MeV LCPs and IMFs.  The GIC-Si telescopes provided
Z resolution for Z = 1 -- 16 fragments in the energy interval from
E/A $\sim$ 0.8-8.0 MeV.  Due to the GIC energy resolution, mass
identification was not possible in the GIC/Si pair.  The final element
of each telescope was a 28 mm-thick CsI(T$\ell$) crystal, read out by a
photodiode.  These crystals provided an energy acceptance of 1 MeV
$\lesssim$ E/A $\leq$ 92 MeV. The Si-CsI(Tl) pair provided Z and
A identification for 8 MeV $\leq$ E/A $\leq$ 92 MeV particles and
``grey particle'' detection for fast protons and pions up to $\sim$ 350 MeV.
The ISiS telescopes did not detect neutrons or Z-identify heavy
residues and fission fragments.

Detector signals were shaped and amplified by means of charge-sensing
preamplifier/linear shaper NIM units, with gains custom-designed for
each detector type.  Analog signals were digitized by 12-bit
16-channel peak-sensing ADCs and fast signal discrimination and
multiplicity sensing was accomplished with 16-channel discriminators
and time-to-digital converters.  Voltages for each detector type were
supplied by in-house-designed, computer-controlled bias supply units.

The event hardware trigger was generated from the fast outputs of the
discriminators and required a minimum of three detected particles.  This
criterion was imposed by the high backgrounds associated with
synchrotrom accelerators and biased the data against low-excitation
energy events (E*\/A $\lesssim$ 1 MeV). In software, only events 
with three thermal particles, one of which
with Z $\geq$ 2, were accepted.  More complete details for the ISiS
array can be found in \cite{Kw95a}.

Three multifragmentation campaigns were carried out with ISiS: (1)
E228 with 1.8 -4.8 GeV $^3$He ions at the Laboratoire National Saturne
in Saclay, France; (2) E900 at the Brookhaven AGS accelerator with
5.0-14.6 GeV/c proton and $\pi^-$ beams; and (3) E900a at AGS with
tagged 8.0 GeV/c antiproton and $\pi^-$ beams.  The number of events with each
beam is summarized in Table 1.

In the following sections we present the results obtained with ISiS. We
first examine the collision dynamics, then the thermal observables,
and finally the thermodynamics and scaling-law behavior of the data.

\begin{table}
\caption{Number of events analyzed for each system (in parentheses).  
Event acceptance requires
at least three thermal charged particles in silicon detectors, one of which
must have Z $\geq$ 2.}
\label{table1}
\begin{tabular}{|c|c|c|c|c|c|}
\hline
\underline{Beam} & \underline{Target} & &
\multispan{2}\underline{Energy/Momentum}& \\


$\pi^{-}$ & $^{197}$Au & 5.0 GeV/c & 8.0 GeV/c$^*$ & 8.2 GeV/c & 9.2 GeV/c \\
& & (1.0 $\times$ 10$^6$) & (2.5 $\times$ 10$^6$) & (2.4 $\times$ 10$^6$)
 & (1.4 $\times$ 10$^6$) \\

p  & $^{197}$Au & 6.2 GeV/c & 9.2 GeV/c & 12.8 GeV/c & 14.6 GeV/c \\
& & (2.4 $\times$ 10$^5$) & (1.7 $\times$ 10$^6$ )& (1.4 $\times$ 10$^6$)
 & (1.1 $\times$ 10$^6$) \\

$\bar{p}$ & $^{197}$Au &  & 8.0 GeV/c$^*$ & & \\
&  &  &  (5.5 $\times$ 10$^4$) & & \\

$^3$He  &  $^{nat}$Ag & 1.8 GeV & 3.6 GeV & 4.8 GeV  & \\
&  & (4.9 $\times$ 10$^6$) & (3.0 $\times$ 10$^6$) & (1.9 $\times$ 10$^6$)& \\

 &  $^{197}$Au & 1.8 GeV &  & 4.8 GeV & \\
& & (4.0 $\times$ 10$^5$) & & (2.9 $\times$ 10$^6$) & \\
$^*$tagged beam & & & & & \\
\hline
\end{tabular}
\end{table}

\section{Reaction Dynamics}

\subsection{Excitation Energy Deposition}

In GeV light-ion-induced reactions the dissipation of radial beam
energy into internal excitation of the target-like residue proceeds
through a complex fast cascade of nucleon-nucleon collisions.  This
mechanism is abetted by the excitation of $\Delta$ and higher
resonances, followed by the reabsorbtion of some of the decay pions
\cite{Cu81,Wo83,Ya79,To90,Wa96}.  For antiproton beams, excitation-energy
deposition is further enhanced by the reabsorbtion of some fraction of
the 4-5 annihilation pions \cite{St84}.  Although the energy
dissipation process is relatively inefficient, the cascade step is
capable of imparting up to $\sim$ 2 GeV of excitation energy in heavy
nuclei.  Equally important,
randomized/equilibrated residues are produced with a continuous
distribution of excitation energies, essentially providing nearly a complete
excitation function in a single reaction.

\begin{figure}
\vspace{10mm}
\centerline{\psfig{file=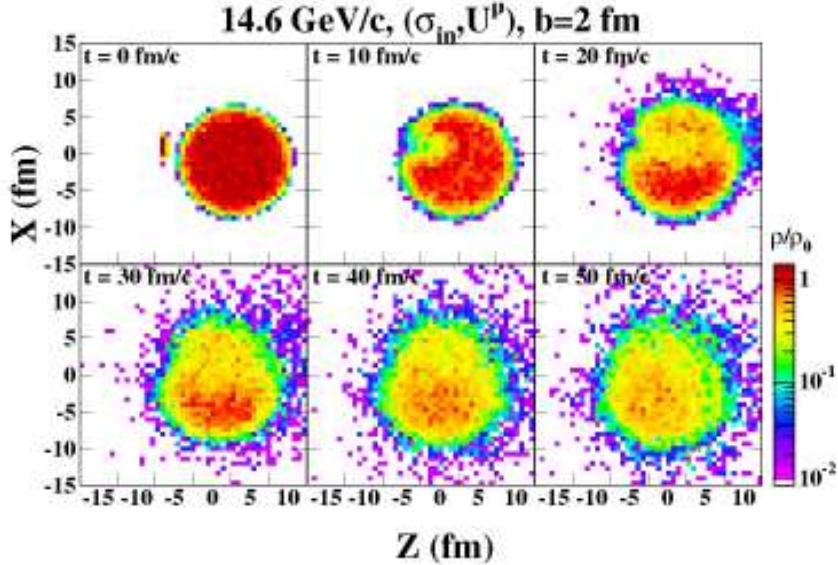,width=4.5in}}
\caption{Nuclear density within the reaction plane XZ around
the position coordinate Y=0, where Z is along the beam axis at
different times in the p + $^{197}$Au reaction at 14.6 GeV/c. Calculation 
is for b=2 fm, with the $(\sigma_{in}, U^p)$ transport simulations 
discussed in Sec. 3.2.\cite{Tu04}}
\label{Fig4}
\end{figure}

Fig. \ref{Fig4} depicts the predictions of a BUU (Boltzmann-Uehling-Uhlenbeck)
calculation 
\cite{Da91,Tu04} that traces the
time and density evolution of a central (b = 2.0 fm) collision
between a 14.6 GeV/c proton and a $^{197}$Au nucleus.  Initially, a
local density depletion develops along the projectile trajectory as
forward-focused nucleons are ejected on a fast time scale.  After
about 30 fm/c the nuclear matter density becomes more uniform,
indicating a random distribution of nucleons.  However, the average
density is observed to be significantly lower than that of the
original target, creating a hot, dilute nucleus.  As the
reaction time increases, the angular distribution of the emitted
particles becomes more isotropic.  At this point the distinction
between low-energy cascade nucleons (nonequilibrium) and evaporative
nucleons becomes blurred.  Also, it is significant to notice that the
heavy residue trajectory has a component transverse to the beam direction.

\begin{figure}
\vspace{22mm}
\centerline{\psfig{file=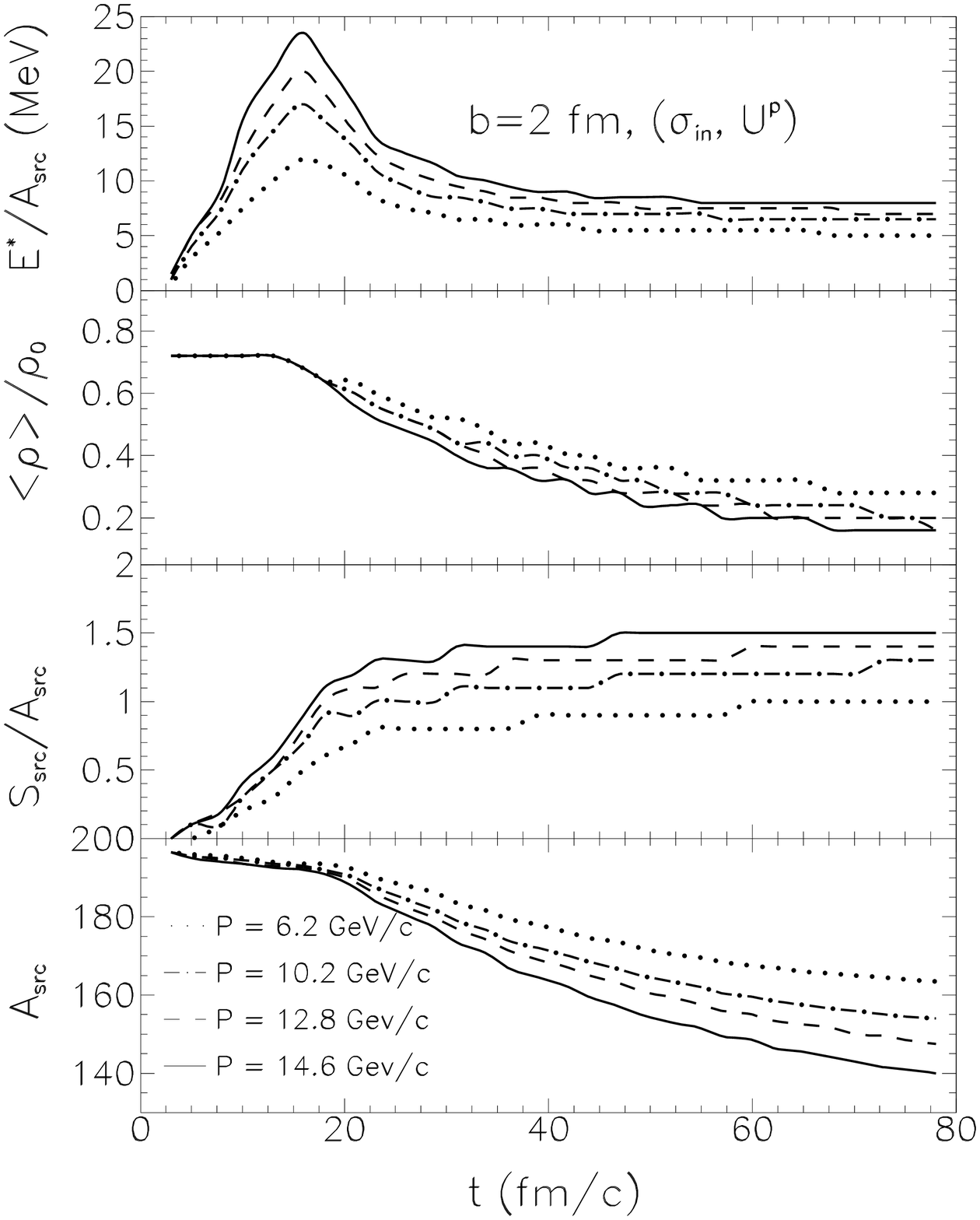,width=3.5in}}
\caption{Excitation energy per nucleon of the source, mean density,
entropy per nucleon and residue mass number, as a function of time in the 
p + Au reaction at b=2 fm and different incident momenta, from the
($\sigma_{in}, U^p$) transport simulations \cite{Tu04}, 
discussed in Sec. 3.2.}
\label{Fig5}
\end{figure}

In Fig.\ \ref{Fig5} the effect of entrance-channel beam momentum on
excitation energy per nucleon (E*/A), average density
($<\rho$$>$/$\rho_0$), entropy per nucleon (S/A) and residue mass (A)
is explored as a function of time.  Calculations are for an impact
parameter b = 2.0 fm for the p + $^{197}$Au reaction at momenta of
6.2, 10.2, 12.8 and 14.6 GeV/c.  It is observed that the excitation
energy and entropy per nucleon increase with beam momenta, while the
source mass and average density decrease.  At long reaction times
there is little difference in E*/A and $<\rho$$>$/$\rho_o$ indicating
a saturation in these variables.  Also, in all cases the entropy per
nucleon remains nearly constant beyond 30-40 fm/c, consistent with the
existence of a randomized system.  The most significant dependence on
increasing beam momentum is the systematic decrease in residue mass,
or in terms of the emitted particles, a greater contribution to the
cascade/nonequilibrium yield.

\begin{figure}
\vspace{25mm}
\centerline{\psfig{file=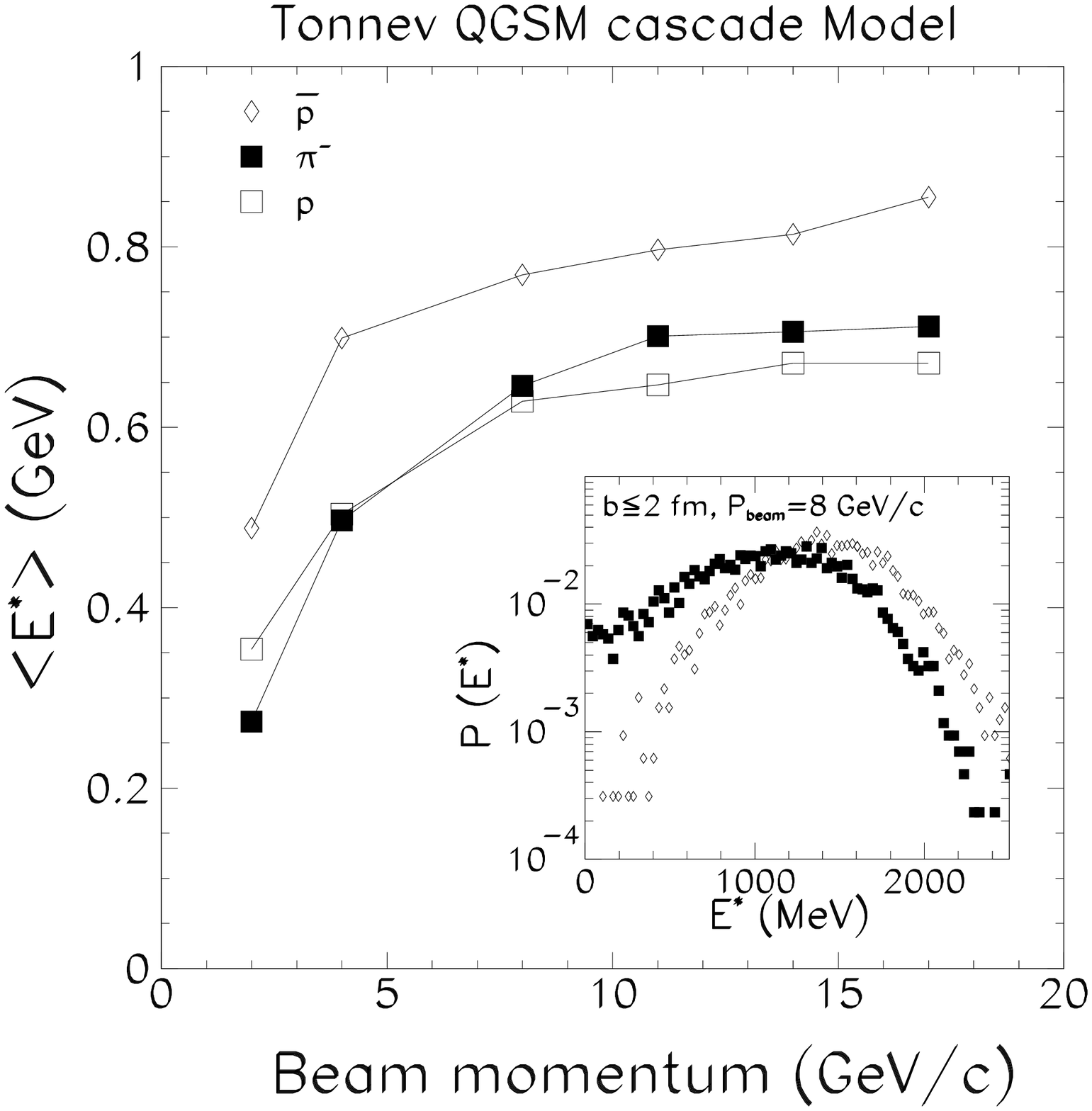,width=3.5in}}
\caption{Intranuclear cascade predictions \cite{To90} of the average
excitation energy for events with E* $>$ 50 MeV are shown as a
function of beam momentum for p, $\pi^-$ and $\bar{p}$ beams incident on
$^{197}$Au.  Inset compares the excitation energy probability
distributions for 8 GeV/c $\pi^-$ and $\bar{p}$ beams}
\label{Fig6}
\end{figure}

All p + A reaction dynamics calculations predict that the excitation
energy probability decreases exponentially with increasing excitation
energy.  For this reason, it is of interest to investigate the
relative efficiency of different hadron probes in depositing
excitation energy.  Fig. \ \ref{Fig6} shows results of an intranuclear
cascade calculation \cite{To90,Le99,Be99} for the average excitation
energy (E* $>$ 50 MeV) as a function of beam momentum for proton,
negative pion and antiproton beams.  For the p and $\pi^-$ cases there
is little difference, since the cascades follow similar paths.  On the
other hand, the annihilation pions from the $\bar{p}$ interaction
enhance $<$E*$>$ significantly.  The inset in Fig. \ref{Fig6}
demonstrates that the excitation-energy distribution for antiprotons is
expected to extend to higher values than for protons and pions.

In comparing the predictions of the dynamics codes with experimental
data, the primary quantities of
interest are the energy deposited in the statistical residue E* and
its mass A.  To do so, however, requires separation of
cascade/preequilibrium emissions from those associated with the
equilibrium-like heavy residue prior to event reconstruction.  This
calorimetry procedure is described in Sec. 4.1.

\begin{figure}
\vspace{20mm}
\centerline{\psfig{file=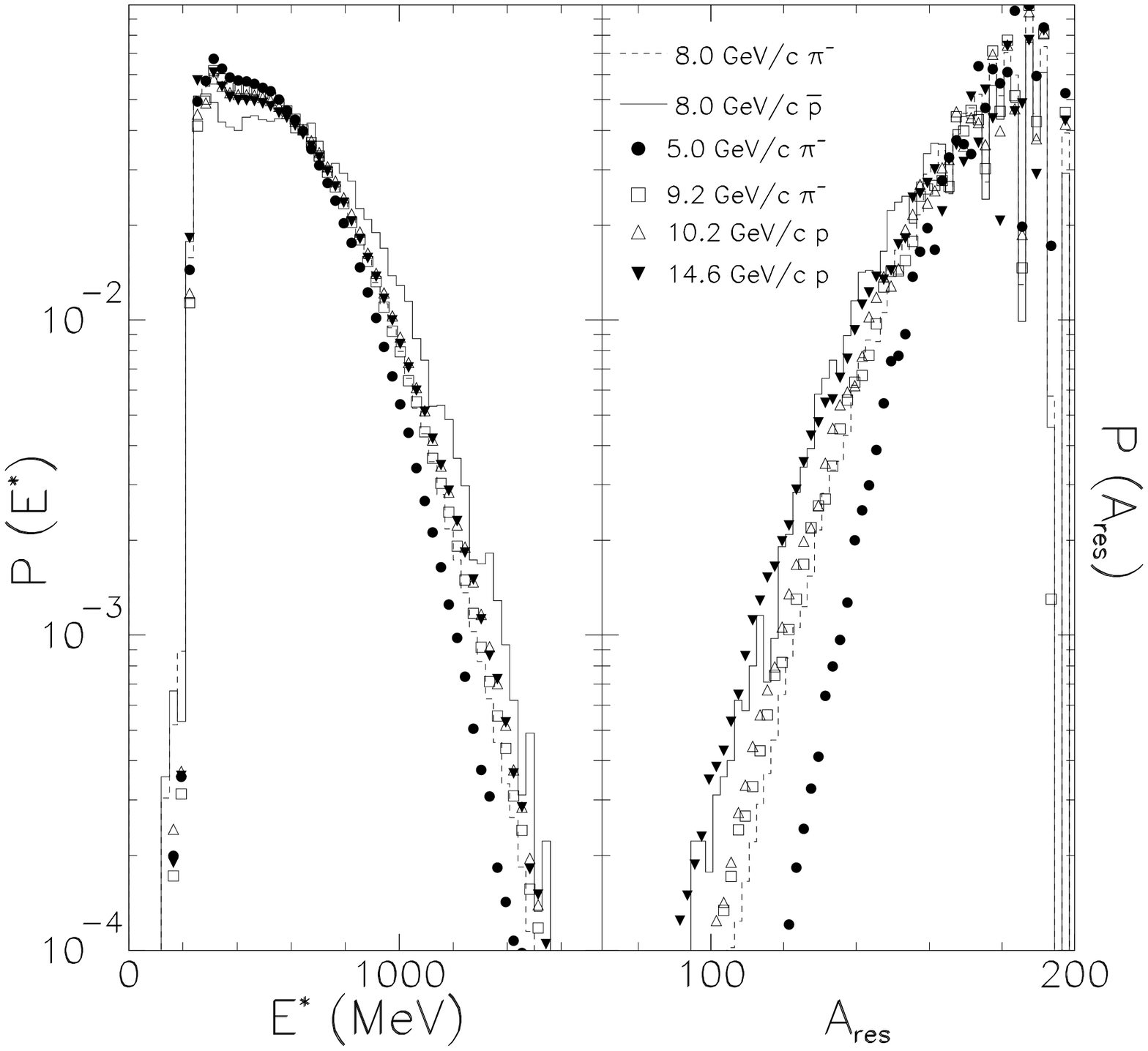,width=4in}}
\caption{Measured excitation energy (left frame) and residue mass (right frame) probabilities for several of the systems studied in this work, as
indicated on the figure.  $\sum P(E^*) = 1$. Data for 6.2 and
12.8 GeV are not shown but are consistent.  Values $<$ 250 MeV are
uncertain because of missing neutrons in the calorimetry procedure.}
\label{Fig7}
\end{figure}

In Fig. \ \ref{Fig7} we show the reconstructed probability
distributions for excitation energy and residue mass for several
systems studied in this work.  The reconstructed E* distributions
shown in the left panel of Fig. \ref{Fig7} demonstrate that the
largest population of high excitation-energy events is achieved with
the 8.0 GeV/c $\bar{p}$ beam and the lowest with the 5.0 GeV $\pi^-$
beam.  Thus, the data are qualitatively in agreement with the INC
calculations (intranuclear cascade) shown in Fig.\ \ref{Fig6},
although the calculations extend to somewhat higher energies than the
data.  The residue mass distributions in the right panel of Fig. \
\ref{Fig7} show a different pattern.  In this case the 14.6 GeV/c
proton beam produces the lightest residues and the 5.0 GeV/c $\pi-$
beam the heaviest, a result relatively well reproduced by the
calculations.  This mass dependence on beam momentum can be understood
as a consequence of the fast cascade, which produces an increasing
number of fast knock-out particles as the beam momentum increases
\cite{Cu81,Wa96}.  This process produces the saturation in excitation
energy observed for hadrons with momenta greater than $\sim$ 8 GeV/c.
That is, the increase in total beam energy available for E* deposition
is counter-balanced by the the loss of energy due to mass loss
$\Delta$A during the fast cascade.

\begin{figure}
\vspace{20mm}
\centerline{\psfig{file=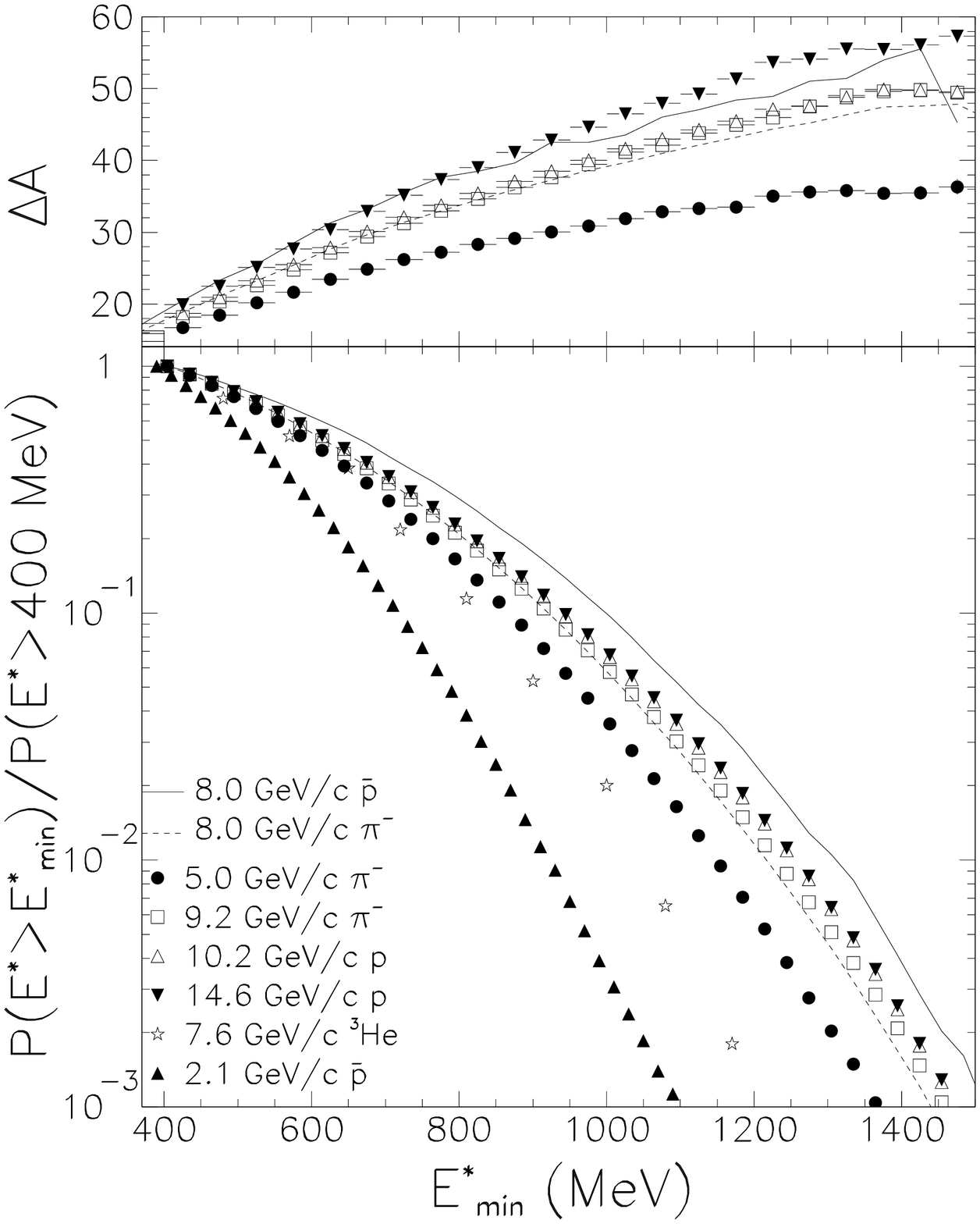,width=3.5in}}
\caption{Bottom: the probability for observing events with excitation
energy greater than E* $\geq$ 400 MeV relative to the probability for
events with E* = 400 MeV.  Systems are indicated on the figure.  Top:
average mass loss $\Delta$A in the fast cascade as a function of
excitation energy.  Systems are defined in bottom frame.}
\label{Fig8}
\end{figure}

The relative effectiveness of various beams in depositing high
excitation energies (Fig. \ref{Fig7}) is emphasized in the bottom
panel of Fig. \ \ref{Fig8}.  Included here are comparable data from
the 4.8 GeV $^3$He + $^{197}$Au reaction \cite{Be01} and from the 1.2
GeV $\bar{p}$ + $^{197}$Au reaction \cite{Go96}.  In order to
emphazise the probability for forming highly excited systems, all data
are normalized to probability P(E*) = 1 at E* = 400 MeV.

Fig. \ref{Fig8} confirms that the 8.0 GeV/c $\bar{p}$ beam produces a
significant enhancement of high excitation energy events, particularly
in the multifragmentation region above E* $\gtrsim$ 800 MeV.  This
figure further supports the predictions of the cascade code; i.e., enhanced
E* with $\bar{p}$ beams, saturation above $\sim$ 8 GeV/c
beam momentum, and exponentially-decreasing probabilities for high
E*/A values.  The behavior of the $^3$He beam can be understood as due
to its average beam momentum of 2.6 GeV/c per nucleon.

\begin{figure}
\vspace{20mm}
\centerline{\psfig{file=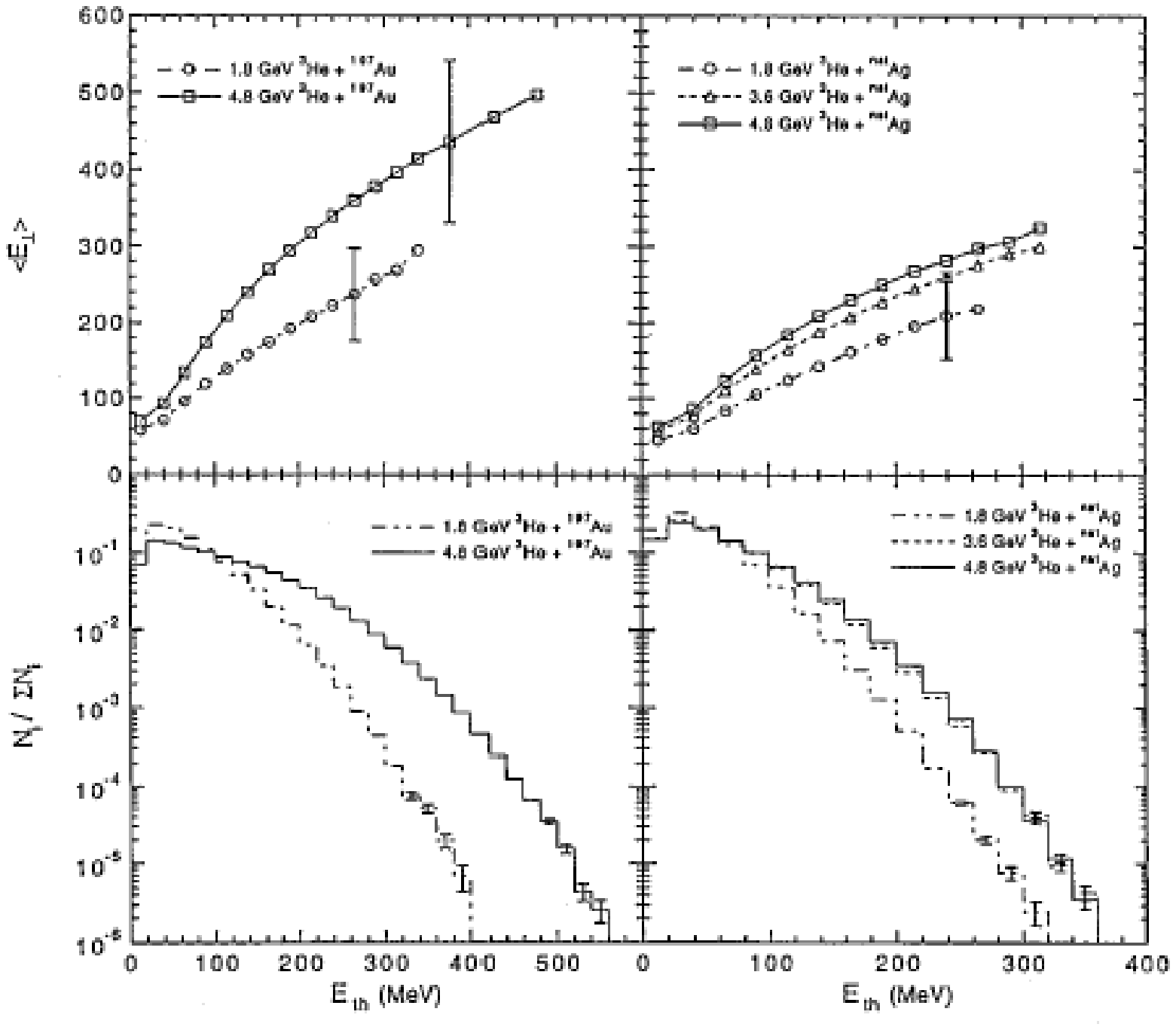,width=4.9in}}
\caption{Lower frames: distributions of observed total thermalized
energy per event for $^3$He + $^{197}$Au(left) and $^3$He +
$^{nat}$Ag (right); upper frames: correlation between total
thermalized energy and transverse energy.  Error bars indicate
standard deviations of distribution widths $(\pm \sigma)$ and are
representative of data.  Systems are defined on figure.}
\label{Fig9}
\end{figure}

Another perspective on the influence of reaction dynamics is provided
by studies with $^3$He beams \cite{Kw98,Mo96}.  Fig. \ref{Fig9} shows the
probability distribution of the total observed thermalized energy and
the correlation with the transverse fragment kinetic energies for the
$^3$He + $^{nat}$Ag and $^{197}$Au reactions.  The thermal energy, which is
strongly correlated with excitation energy, is significantly lower for the lighter $^{nat}$Ag target.
However, when source mass corrections are made, the energy per nucleon
distributions are very similar for both targets \cite{Kw98}.  The
projectile-target effect on the saturation of excitation energy with
beam energy is in good agreement with INC calculations
\cite{Ya79,Mo96} for this lower momentum, complex projectile.  The
lower right-hand frame of Fig. \ref{Fig9} shows that the probability
distributions are identical for the $^3$He + $^{nat}$Ag at 3.6 and 4.8
GeV, indicating the onset of deposition energy saturation near 3.6
GeV.  The slight difference at 3.6 and 4.8 GeV for the transverse
energy correlations is due to the difficulty in removing
nonequilibrium events from this sum, which illustrates the uncertainties involved with the use of transverse energy as a gauge of excitation energy depostion.

\subsection{BUU Simulations}

\begin{figure}
\vspace{25mm}
\centerline{\psfig{file=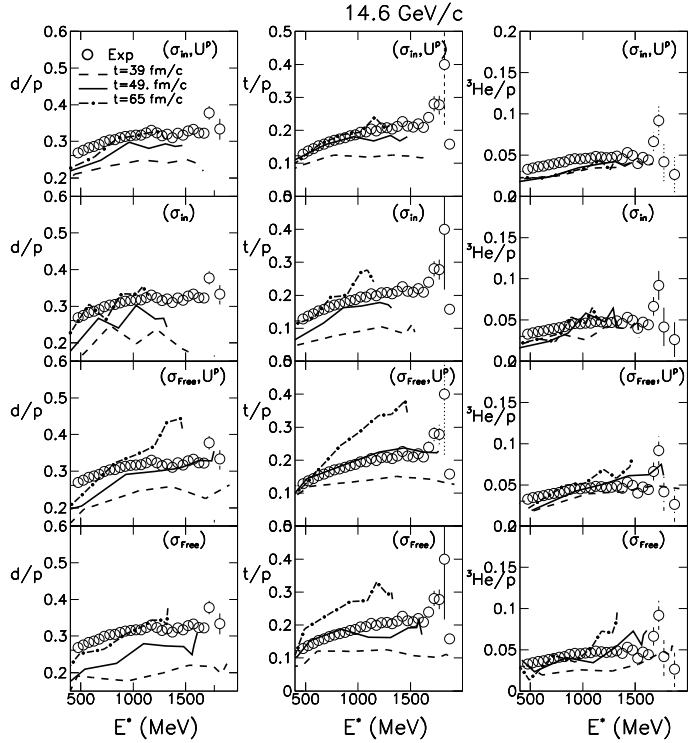,width=4.0in}}
\caption{Normalized yields of nonequilibrium light 
charged particles as a function of
the source excitation energy for the p+Au reaction at 14.6 GeV/c.  Circles
represent data and lines represent filtered ratios from different
versions of the transport calculations, at different times in the
reaction. }
\label{Fig10}
\end{figure}

While the equilibrium-like events are of primary interest for
multifragmentation studies, the nonequilibrium component of the
spectrum is important for understanding the reaction dynamics.  In
order to investigate this facet of the data, predictions of a BUU code
that includes d, t and $^3$He cluster formation have been compared
with cascade/preequilibrium d/p, t/p and $^3$He/p ratios for the p +
$^{197}$Au reaction between 6.2 and 14.6 GeV/c \cite{Da91,Tu04}.  BUU
calculations were performed for various reaction times with and
without a momentum-dependent potential and with both free and
in-medium cross-section options.  Results for the 14.6 GeV/c p +
$^{197}$Au reaction are shown in Fig. \ref{Fig10}.  From examination
of the LCP ratios for all four bombarding energies, the best agreement
with all the data is found when both a momentum-dependent potential
and in-medium cross-sections are employed in BUU code, with a most
probable reaction time of t $\sim$ 65 fm/c.  From comparison with
Fig.\ \ref{Fig5}, entropy considerations suggest that randomization of
the nucleon momenta in the heavy residue occurs after about t $\sim$
30 fm/c, indicating an additional 30-40 fm/c is required to reach a
state of quasiequilibrium, after which the residue undergoes
statistical decay.

Further, the BUU calculation that incorporates clusters, a
momentum-dependent potential, in-medium cross sections and a time
scale of t $\sim$ 65 fm/c is able to describe the excitation-energy
probability distribution and the average source Z and A as a function
of E*/A.  One important aspect of this code is that the inclusion of
clusters serves to enhance energy deposition in the heavy
residue. This feature is missing in other codes designed for transport
calculations with GeV hadron beams.

\subsection{Sideways Peaking}

Finally, the ISiS data demonstrate the important role that reaction
dynamics exert on the statistical decay properties of hot residues.
Earlier inclusive studies \cite{Cu64,Re75,Fo78} showed the existence
of sideways peaking of IMF angular distributions for p +
A reactions above $\sim$ 10 GeV.  This result was interpreted as
possible evidence for dynamical IMF emission during the initial phases of the
cascade, possibly signaling the existence of nuclear shock wave
effects.  Exclusive IMF angular distributions obtained with ISiS support
a more mundane origin for the sideways peaking.

\begin{figure}
\vspace{25mm}
\centerline{\psfig{file=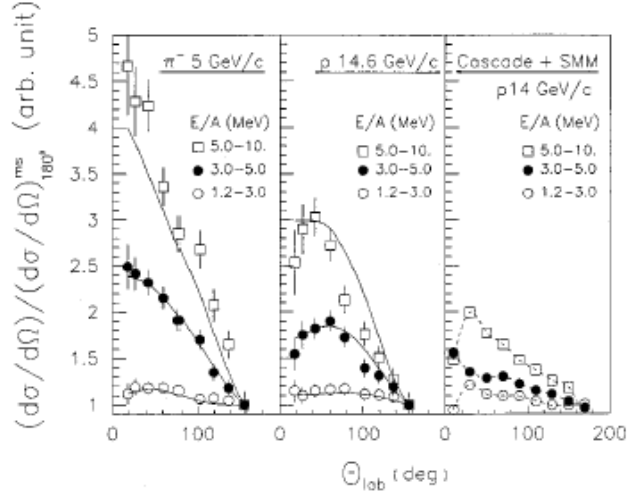,width=3.5in}}
\caption{Dependence of relative angular distributions on IMF kinetic
energy for Z = 5 - 9 fragments formed in events with IMF multiplicity
M $\geq$ 4 for 5.0 GeV/c (left) and 14.6 GeV/c proton (right) beams
incident on $^{197}$Au \cite{Hs99}.  Angular distributions are 
normalized to 1.0 at 160$^{\circ}$.  IMF kinetic-energy bins 
are indicated in the figure.  Solid lines in the left and center panels 
are results of moving-source fits.
Right-hand panel shows prediction of an INC/SMM 
calculation \cite{To90,Bo90} for
the 14 GeV/c p + $^{197}$Au reaction, binned the same as the data.}
\label{Fig11}
\end{figure}

In \cite{Hs99} it is confirmed that the sideways peaking develops only
above beam momenta of 8-10 GeV/c and that degree of peaking increases
as IMF multiplicity and charge increase.  This result indicates that
the peaking is associated with high deposition energy collisions,
where multifragmentation is the major decay mode. In Fig. \ref{Fig11}
relative angular distributions for Z = 5-9 fragments produced in
M$_{IMF} \geq$ 4 events are compared for reactions with 5.0 GeV/c
$\pi^-$ (left panel) and 14.6 GeV/c p (center panel) beams on a
$^{197}$Au target.  Fragment kinetic-energy cuts of E/A = 1.2-3.0,
3.0-5.0 and 5.0-10.0 MeV are imposed on the spectra and all angular
distributions are normalized to unity at 160$^\circ$. (In this regard
it should be stressed that most of the cross section is concentrated
in the kinetic energy bins below 5 MeV.) It is observed that as the
IMF kinetic energy decreases, the angular distributions become
increasingly isotropic for both energies. However, whereas for the
$\pi^-$ beam the angular distributions remain forward-peaked, for the
14.6 GeV/c proton case sideways peaking is observed -- and the maximum
differential cross section evolves to larger angles as the IMF
velocity decreases.

The observation that sideways emission is favored by high beam
momentum, high IMF charge and low IMF kinetic energy suggests a
possible origin in the kinematics of the residue rather than in
dynamical emission.  The diffractive nature of the initial N-N
collision at GeV momenta preferentially produces a secondary nucleon
or N* that recoils 70$^\circ$ - 90$^\circ$ to the beam axis, with the
angle growing as the beam momentum increases.  Subsequent dissipation
during the cascade imparts a transverse velocity component to the
heavy residue.  The net result is that statistical fragment emission
from the residue is focused non-axial to the beam direction -- which
affects the lowest energy IMFs most strongly.  This conjecture is
reinforced by INC/SMM (Statistical Multifragmentation Model)
calculations \cite{To90,Bo90}, shown in the right-hand frame of Fig. \
\ref{Fig11} and is also illustrated in Fig. \ref{Fig4}.  Thus, these
results, coupled with IMF-IMF angular correlations and sphericity and
coplanarity distributions, described in \cite{Fo96}, do not support
arguments for dynamical effects such as shock waves as a source of
sideways-peaking observed in inclusive angular distribution studies.

\section{Statistical Decay:  Multifragmentation}

\subsection{Calorimetry}

In any attempt to describe a system in terms of thermodynamics, a
knowledge of the heat content is fundamental.  For hot nuclei, this
energetic factor is expressed in terms of the excitation energy per
residue nucleon, E*/A.  In this section we examine the procedures for
determining E* and A for the ISiS data \cite{Kw98,Le01,Ru02}.

For each reconstructed event, the excitation energy of the emitting
source is calculated as follows:

\begin{equation}
E^*_{source} = \sum^{M_{cp}}_{i}K_{cp}(i) + M_n < K_n> + E_{\gamma}-Q.
\end{equation}

\noindent Here K$_{cp}$ is the kinetic energy for all thermal charged
particles, M$_n$ is the multiplicity of thermal neutrons with average
kinetic energy $<K_n>$, E$_\gamma$ is the total energy emitted by
gammas, and -Q is the removal energy (the negative of the reaction
Q-value).  Each of these terms requires assumptions, as described in
the following.

\begin{figure}
\vspace{35mm}
\centerline{\psfig{file=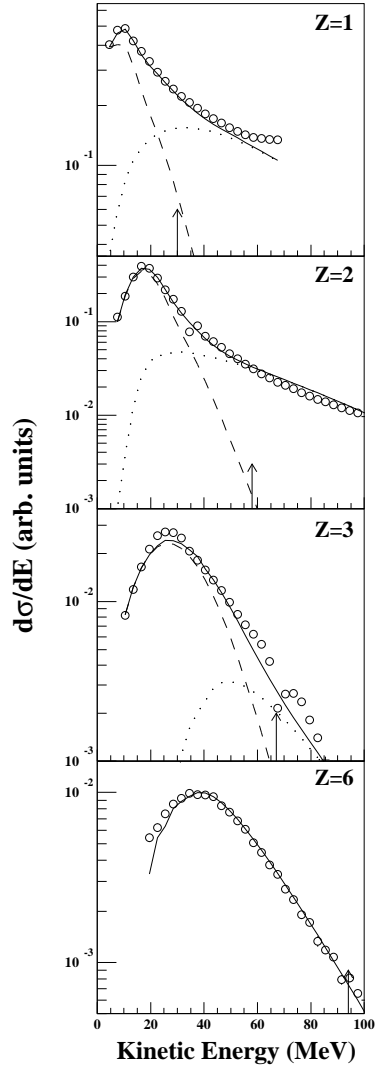,width=2.1in}}
\caption{Angle-integrated kinetic-energy spectra in the laboratory frame for
Z = 1, 2, 3 and 6, as indicated in each panel, emitted in the 8.0 GeV/c $\pi^-$ + $^{197}$Au reaction.  Open points correspond
to data.  Dashed (dotted) lines represent the thermal-like
(nonequilibrated) component of the moving-source fit.  The solid line
is the sum of the two fits.  Upper cutoff energies \cite{Mo96} are
shown by vertical arrows.}
\label{Fig12}
\end{figure}

In calculating the charged particle contribution to E* in Eq.(1),
cascade/preequilibrium emissions prior to thermalization must be
removed from the sum.  Separation on an event-by-event basis is not
fully transparent due to the time evolution of the cooling process.
Fig. \ref{Fig12} presents angle-integrated spectra for Z = 1,2,3 and 6
nuclei from the 8.0 GeV/c $\pi^-$ + $^{197}$Au reaction.  Because
source velocities are low ($\sim$ 0.01 c), kinematic effects are
small.  The principal features of the spectra are a Maxwellian
low-energy component, which we attribute to thermal events, and an
exponentially-decreasing high-energy tail due to nonequilibrium
processes.

\begin{figure}
\vspace{25mm}
\centerline{\psfig{file=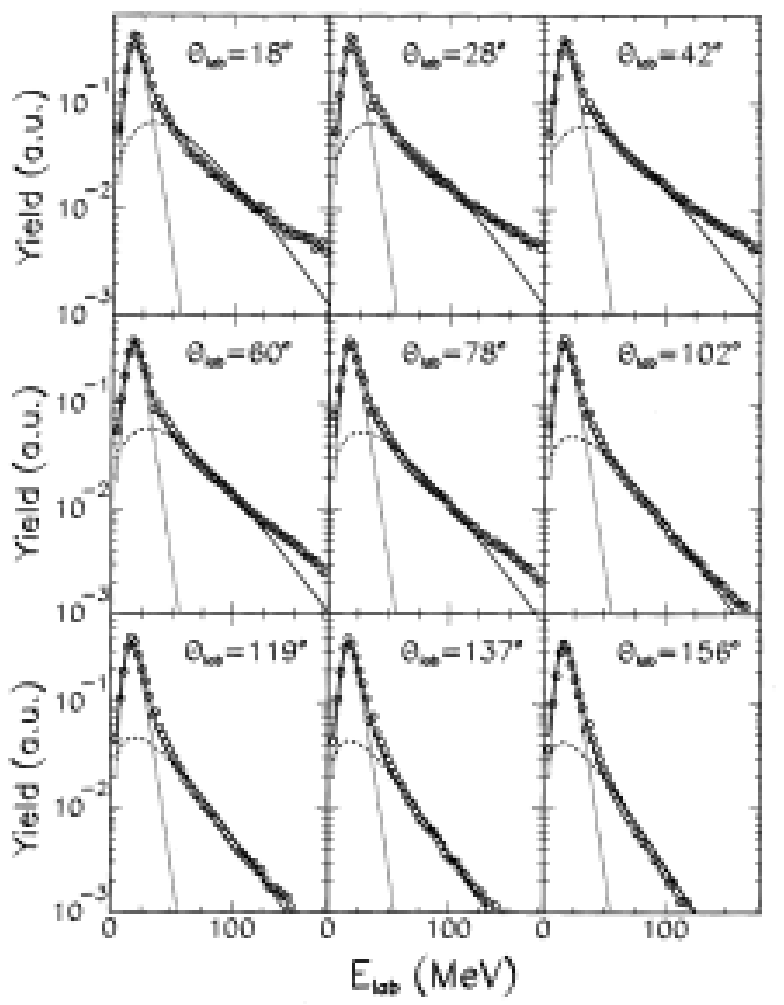,width=3.0in}}
\caption{Angle-dependent spectra for Z = 1.  
Reaction is 8.0 GeV/c $\pi^-$ + $^{197}$Au.}
\label{Fig13}
\end{figure}

The spectra have been decomposed \cite{Mo96,Le01,Ru02} with a
two-component moving-source model \cite {We78}.  In Fig.\ \ref{Fig12}
the thermal source \cite{Mo75,Kw86} is described by the dashed lines,
the nonequilibrium source \cite{We78} by dotted lines and the total by
the solid line.  The nonequilibrium component is most important for
the hydrogen isotopes, but is also a significant fraction of the He
yield.  In the proton case the two-component model is insufficient to
account for the high-energy portion of the spectrum, suggesting
evidence for three components: thermal at low energies, preequilibrium
at intermediate energies, and fast cascade particles at the highest
energies.  Such a picture is consistent with the BUU transport
calculations in Sec. 3.  As seen in Fig. \ref{Fig13}, nonequilibrium
emission persists, even at the most backward angles. Fig. \ref{Fig13}
also illustrates the weak-angular dependence of the spectra.  For
IMFs, the nonequilibrium component decreases in yield as the IMF
charge inreases.  For carbon and heavier IMFs, the angle-integrated
preequilibrium yield is negligible.

Since H and He isotopes constitute most of the charged-particle yield, 
calorimetry requires a systematic procedure for
distinguishing between thermal and nonequilibrium emissions on an
event-by-event basis. This need was a primary motivation for the
moving-source fits to the spectra.  From moving-source analyses of all
the spectra from the 1.8-4.8 GeV $^3$He + $^{197}$Au reaction
\cite{Mo96}, it was concluded that a sharp cutoff approximation gave a
satisfactory account of the fraction of thermal events in the spectra.
The thermal cutoff values were:

\begin{equation}
K_{cp} ( Z = 1) \leq 30 MeV \ \ \ {\rm and}
\end{equation}

\begin{equation}
K_{cp} (Z \geq 2) \leq (9Z + 40) MeV.
\end{equation}

\noindent These cutoffs are indicated by the arrows in
Fig. \ref{Fig12}, and correspond approximately to the break in the
slope of the spectral tails.  This method of determining the thermal
yield was also compared with the integrated yield from the
two-component fits, which led to slightly lower E* values \cite{Le01}.
For the calculation of excitation energy in this work, the cutoff
values of Eqs. (4.2) and (4.3) were employed.  The resulting thermal
yields are isotropic in the center-of-mass frame \cite{Le01}.

\begin{figure}
\vspace{25mm}
\centerline{\psfig{file=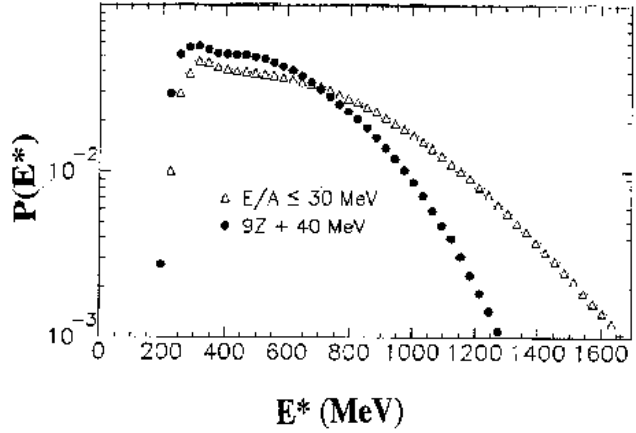,width=3.5in}}
\caption{Excitation-energy distributions for sharp cutoff assumptions of 
Eqs.(4.2) and (4.3) compared with a cutoff value of K$_{cp} <$ 30A MeV.}
\label{Fig14}
\end{figure}

\begin{figure}
\vspace{25mm}
\centerline{\psfig{file=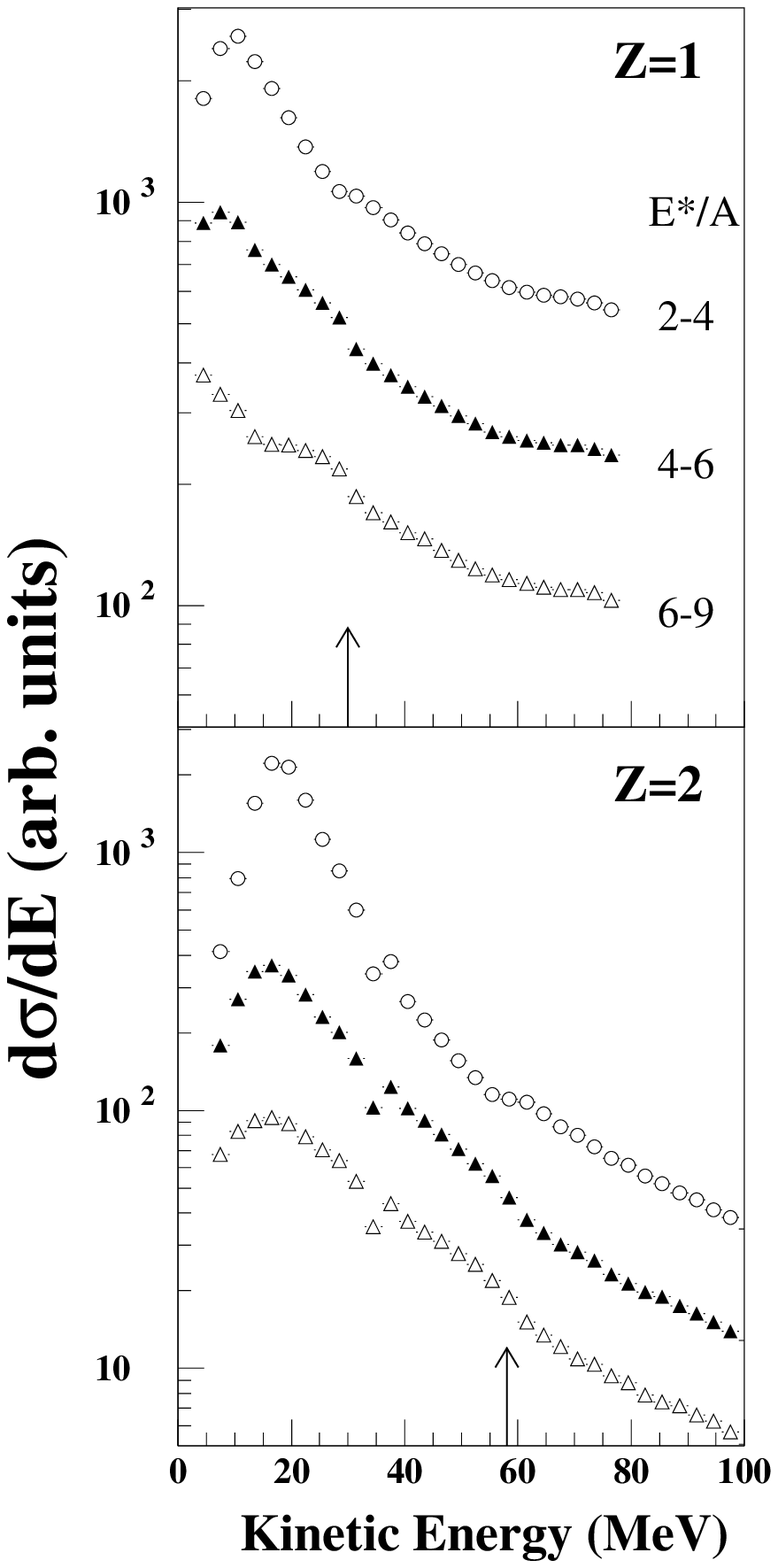,width=2.0in}}
\caption{Angle-integrated kinetic-energy spectra in the laboratory for
Z=1 and Z=2 as indicated in each panel, and for three bins in E*/A
calculated with the cutoff assumptions of Eqs. (4.2) and (4.3) \cite{Mo96}.
The symbols correspond to bins of E*/A=2-4 (open circles), E*/A=4-6 (filled
triangles), and E*/A=6-9 MeV (open triangles).  Reaction is 8.0 GeV/c $\pi^-$ + $^{197}$Au.}
\label{Fig15}
\end{figure}

In contrast the E* analysis of the EOS 1 GeV $^{197}$Au + $^{12}$C data
\cite{Ha96} used a sharp cutoff assumption of E/A $\leq$ 30 MeV for 
all fragments. This
approach yields significantly higher E* values, as shown in
Fig. \ref{Fig14}, largely due to the inclusion of preequilibrium He
ions in the sum of Eq. (1).  When this difference is taken into
account, the ISiS and EOS experiments are in good agreement in those
areas where they overlap.  While the EOS calibration may lead to high
E* values, the ISiS sharp cutoff approximation may underestimate E*
for high excitation energies.  This problem is illustrated in
Fig. \ref{Fig15} where the LCP kinetic energy spectra are plotted for
several E*/A bins.  The sharp cutoff distinction is clear at low
excitation energies, but the two components blend into one as E*/A
increases, blurring the separation.

\begin{figure}
\vspace{25mm}
\centerline{\psfig{file=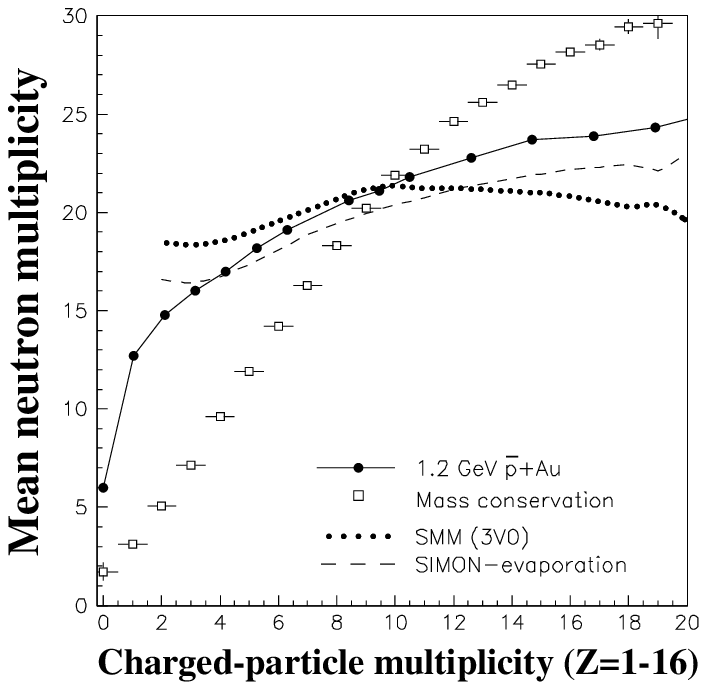,width=2.9in}}
\caption{Relation between the mean neutron multiplicity and the
charged-particle multiplicity.  Solid line corresponds to data points
reported for LEAR data by Ref. \cite{Go96}; dotted line shows the SMM
calculation, and the SIMON evaporation result is given by the dashed line.  The
open squares give the mean multiplicities estimated with the mass
conservation assumption \cite{Ma97,Be95}.}
\label{Fig16}
\end{figure}

The second major uncertainty in determining E*/A for the ISiS data is the
unmeasured neutron contribution to both the thermal sum in Eq. (1) and
the cascade/preequilibrium multiplicity as it affects the source mass
\cite{Kw98,Le01}.  In order to estimate the thermal-like neutron
component, we have normalized the neutron charged-particle
correlations reported by \cite{Go96} to the ISiS charged-particle
results.  The measured correlations, shown in Fig. \ref{Fig16}, are
reasonably well described for charged-particle multiplicity M $\geq$ 4
by model simulations \cite{Bo90,Du92} and show the same qualitative
behavior as has been observed in heavy-ion reactions \cite{To95}.  A
mass-balance procedure \cite {Ma97} does not work well for the ISiS
data.  The rapid rise in neutron multiplicity at low energies, where
charged-particle multiplicities are low, makes the ISiS E* values
increasingly uncertain below E* $\lesssim$ 200 MeV.

Several assumptions have been employed in order to estimate the
average neutron kinetic energy as a function of E*/A.  For the 4.8 GeV $^3$He
reactions $<K_n>$ was estimated from Coulomb-corrected proton spectra
and then iterated to obtain a consistent value $<K_n> = 2T_{th}$,
where $T_{th}=(E*/a)^{1/2}$ and a = A/11 MeV$^{-1}$ \cite{Kw98}.  For
the hadron-induced reactions several relationships were explored,
including Fermi-gas and  It should be noted that the IMF yield is actually
largest for the E/A = 1.2-3.0 MeV bin and smallest for the 5.0-10.0
MeV bin. Maxwell-gas assumptions with level density
parameters a = A/8 MeV$^{-1}$ and a = A/13 MeV$^{-1}$
\cite{Le01}. Comparisons were also made with SMM \cite{Bo90} 
and SIMON \cite{Du92} evaporation codes.  Based on this
analysis, the SMM predictions were used as a conservative estimate of
the neutron kinetic energy contribution to E*.  Eq.(4.1) is then
iterated to obtain self-consistency.  This procedure produced a
somewhat lower neutron kinetic energy contribution than in \cite{Kw98}.
For both the neutron multiplicities and kinetic energies, the use of
averages leads to loss of fluctuation information in the final excitation
energies.

The sharp-cutoff and neutron assumptions, along with the detector
geometry, are included in the detector filter.  The minor contribution
to Eq. (4.1) from gamma emission is assumed to be E$\gamma$ =
2(M$_{cp} + M_n$) MeV in \cite{Kw98} and E$\gamma$ = M(Z $\geq$ 3) MeV
\cite{Le01}.  In order to calculate the removal energy (-Q), the
charge and mass of the source must be reconstructed.  The source
charge is determined from

\begin{equation}
Z_{source} = Z_{tgt} - \sum^{Mneq}_i Z_i(neq),
\end{equation}

and the mass from

\begin{equation}
A_{source} = A_{tgt} - \sum^{Mneq}_i Z_i(neq) - <M_n(neq)>,
\end{equation}

\noindent where Z$_{tgt}$ and A$_{tgt}$ are the charge and mass of the
target, Z$_i$(neq) is the charge of the measured
cascade/preequilibrium particles, and the nonequilibrium neutron
multiplicity is related to the nonequilibrium proton multiplicity by
$<M_n^{(neq)}> = 1.93 M_p^{(neq)}$.  The assumption for
$<M_n^{(neq)}>$ is consistent with BUU calculations \cite{Tu04} and
experimental results \cite{Eg00}, and is intermediate between the A/Z
of the target and experimental systematics \cite{Pol95}.  The
unmeasured IMF mass is based on the isotopically-resolved data of
\cite{Gr84}. In the calculation of E*/A the resultant Q values and
neutron multiplicities partially offset one another in Eq. (4.1);
e.g., if the neutron term is over-estimated, then the removal energy
is reduced, and vice versa.

Several second-order corrections have been investigated and found to
have no significant effects, among them:  the source velocity ($\sim$
0.01c), source emission angle \cite{Hsi98}, and detector threshold
effects.  Since ISiS does not measure heavy residues, the additional
assumption is made that all missing mass and charge are contained in a
single residue. The residue mass distribution obtained in this way is
in good agreement with measured results from \cite{Ha96}.  The
consistency of the ISiS calorimetry filter has been tested with SMM
\cite{Bo90} and SIMON \cite{Du92} calculations that use the measured
source mass, charge and excitation energy as inputs to the codes
\cite{Le01}.  One final consideration is the effect of the
exponential decrease in the E*/A probability with increasing E*/A,
which serves to decrease the excitation energy relative to the bin average. 
This effect is most significant for the highest 
excitation energy bins when the measured
distribution is deconvoluted.

\begin{figure}
\vspace{25mm}
\centerline{\psfig{file=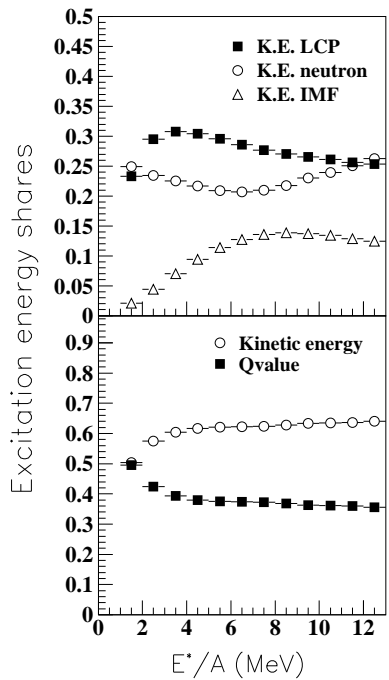,width=2.5in}}
\caption{Relative share of excitation energy for various components of
the reconstruction procedure of Eq.(4.1) as a function of
E*/A for the 8.0 GeV $\pi^-$ + $^{197}$Au reaction. 
Top frame: light-charged particle kinetic energy (solid
squares), neutron kinetic energy (open circles), and IMF kinetic
energy (open triangles).  Bottom frame: total particle kinetic energy
(open circles) and Q values (solid squares).}
\label{Fig17}
\end{figure}

The relative fractions of the excitation energy for the LCP, IMF and
neutron kinetic energy contributions to Eq. (4.1) are plotted in the top
frame of Fig. \ref{Fig17}.  Over the full range of E*/A the LCP
fraction ranges from 25-30\%, and that for neutrons from 20-25\%,
accounting for over half of the total E*. The IMF kinetic energy
fraction is relatively small, ranging from negligible values at low
excitation energies to a near-constant value of $\sim$ 10-12\% above
E*/A $\approx$ 6 MeV.  In the bottom frame of Fig. \ref{Fig17} the excitation
energy fraction due to removal energy (-Q) is compared with the
fraction for total kinetic energy.  Except for low E*/A, where the
calorimetry is most uncertain, the total kinetic energy sum is a
near-constant factor of two greater than the removal energy.

Overall we estimate that the values chosen for the ISiS E*/A data and
associated assumptions could reasonably be lowered by $\sim$ 5\% or
increased by $\sim$ 15\%.  In the following sections we examine the
data as a function of the calorimetric E*/A procedure described in this
section.

\subsection{Thermal Observables}

\begin{figure}
\vspace{25mm}
\centerline{\psfig{file=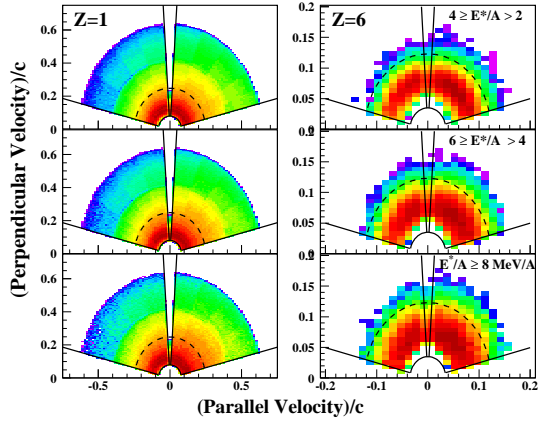,width=3.0in}}
\caption{Contour plot of longitudinal v$_\parallel$
vs. transverse v$_\perp$ velocity of hydrogen (left) and carbon
(right) fragments from the 8.0-MeV/c $\pi^-$ + $^{197}$Au reaction for
several bins in E*/A.  Solid lines indicate geometrical acceptance of
the ISiS array; dashed line gives the thermal cutoff velocity
\cite{Le01}, not corrected for source velocity.}
\label{Fig18}
\end{figure}

Of the many experimental signals for thermal behavior in hot nuclei,
the most transparent are found in the evolution of the spectra as a
function of excitation energy.  Two fundamental tests that a
statistically-decaying system must face are: (1) is particle emission
isotropic? and (2) are the kinetic energy spectra Maxwellian?  In
Fig. \ref{Fig18} invariant cross sections (parallel v$_\parallel$
versus perpendicular v$_\perp$) velocity components are shown as a
function of excitation-energy bins for hydrogen and carbon ions
measured for the 8.0 GeV/c $\pi^-$ + $^{197}$Au reaction.  For
energetic hydrogen ions (v $\gtrsim$ 0.2c) at all E*/A one observes a
spray of forward-emitted particles that originates in
cascade/preequilibrium processes. The low-energy part of these plots
is nearly symmetric about zero velocity, indicating emission from a
randomized source moving with an average velocity of $\sim$ 0.01c.
The ISiS acceptance for thermal-like particles is defined by the
dashed line in Fig. \ref{Fig18}, i.e., the sharp cutoff assumption of
\cite{Mo96}.  The isotropy of the projected angular distributions for
thermal LCP and IMF emission, as well as the relative insensivity to
excitation energy, has been demonstrated in \cite{Fo96,Le01}.  In
addition, an event-shape analysis for the thermal IMFs from the 4.8
GeV $^3$He + $^{197}$Au reaction is found to be consistent with the
existence of a randomized system that disintegrates on a very short
time scale \cite{Fo96}.

\subsubsection{Fragment Spectra}

\begin{figure}
\vspace{25mm}
\centerline{\psfig{file=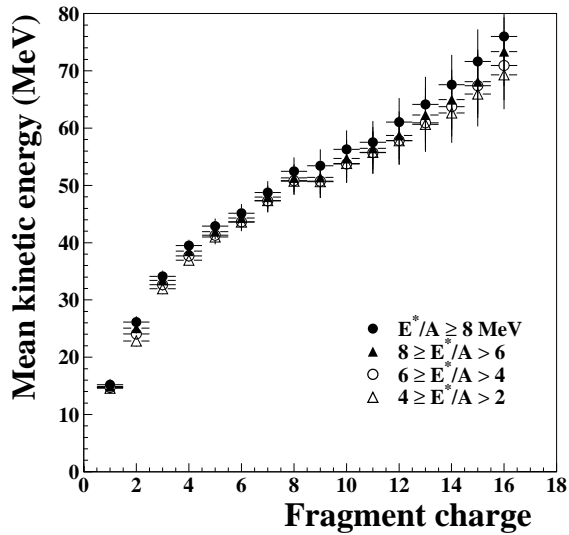,width=3.0in}}
\caption{Fragment mean kinetic energy as a function of IMF charge
calculated in the source frame for four bins of excitation energy, as
indicated on the figure.  Data are from the 8.0-GeV/c $\pi^-$-induced
reaction on $^{197}$Au.}
\label{Fig19}
\end{figure}

The Maxwellian character of the LCP and IMF kinetic-energy spectra is
illustrated in Figs. \ref{Fig12}, \ref{Fig13}, and \ref{Fig15}.
Figures \ref{Fig12}, \ref{Fig13} and \ref{Fig18} show the systematic
Coulomb shift in the spectral peaks due to kinematic behavior and
fragment charge.  Figure \ref{Fig12} reveals two opposing trends that
become apparent when the LCP spectra are gated on E*/A.  First, the
thermal slopes become flatter as E*/A increases, the expected result
of the increasing temperatures.  In contrast, instead of showing the
expected temperature-dependent increase in the spectral (Coulomb) peak
energies for a system at normal nuclear density, the spectral peak
energies decrease.  The net effect of these opposing effects is that
the average mean kinetic energy for thermal particles is essentially
independent of excitation energy, as shown in Fig. \ref{Fig19}.  This
figure also shows the expected increase in the average kinetic energy
as a function of fragment charge.  Within the context of Figs. 15 and
\ref{Fig19}, the observed fragment mean kinetic energies can be
interpreted in terms of a compensation between two competing factors:
an increase in temperature offset by a decrease in the source density
as the excitation energy increases.  This behavior is explored in
greater detail in Sec. 4.3.

\subsubsection{Multiplicities}

\begin{figure}
\vspace{25mm}
\centerline{\psfig{file=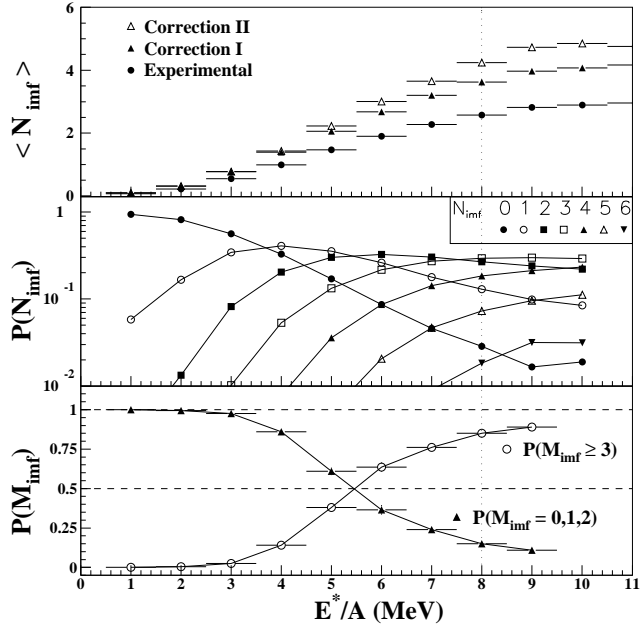,width=3.5in}}
\caption{Top: average number of IMFs for the observed yield (closed circles), 
the yield corrected for geometry (solid triangles),  and the yield corrected 
for both geometry and
fragment energy thresholds (open triangles) as a function of E*/A for the
8 GeV/c $\pi^-$ + $^{197}$Au reaction.  Middle: probability for given
number of detected IMFs as a function of E*/A.  Bottom: probability
for IMF multiplicity M $\geq$ 3 (circles) and M$<$3 (triangles).}
\label{Fig20}
\end{figure}

An important signature of multifragmentation, and its possible
relation to a nuclear liquid-gas phase transition
\cite{Bon85,Fr90,Gr90}, is the multiplicity of IMFs in an event.
The models predict that above E*/A $\sim$ 4-5 MeV, multiple IMF
emission should appear.  Figure \ref{Fig20} examines this prediction,
presenting the average IMF multiplicity (top frame), emission 
probabilities for a fixed multiplicity (middle frame) and the probability
for emitting three or more IMFs relative to two or less (bottom
frame).  The unmeasured heavy residue is not included in these
probabilities and M is derived from a Monte Carlo reconstruction of
the measured fragment multiplicity N that accounts for detector
geometry and thresholds. 

In the top frame the IMF multiplicity is shown to increase
monotonically, with no apparent deviation near E*/A $\approx$ 5 MeV.
However, when the averages are decomposed into specific probabilities
(middle frame), it is observed that as E*/A increases, the probability
for increasing N$_{IMF}$ opens up systematically with increasing
excitation energy.  The probability for emitting three or more IMFs
(the classical definition of multifragmentation \cite{Bon85}) is seen
to grow rapidly near E*/A $\sim$ 4 MeV, so that above E*/A $\sim$ 5
MeV multifragmentation becomes the dominant decay channel.  This
growth in IMF multiplicity is accompanied by a corresponding growth in
the thermal LCP multiplicity.

\subsubsection{Charge Distributions}

\begin{figure}
\vspace{25mm}
\centerline{\psfig{file=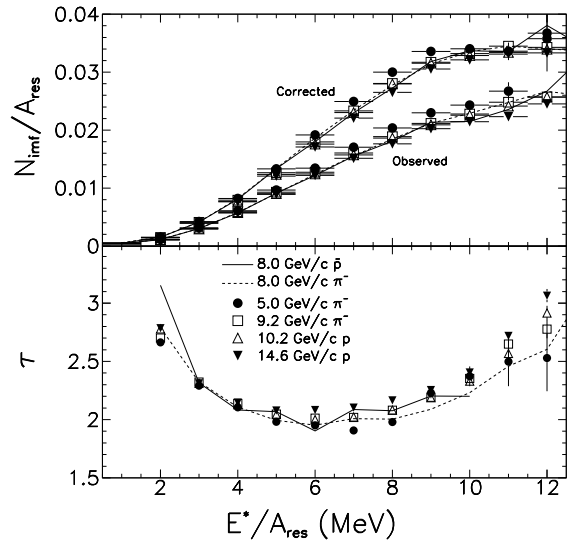,width=3.0in}}
\caption{Top: average ratio of observed and geometry-corrected IMFs
per residue nucleon as a function of E*/A; symbols are defined in the
bottom frame.  Bottom: power law parameters $\tau$ from fits to the
charge distributions as a function of E*/A of the residue. }
\label{Fig21}
\end{figure}

Another important aspect of the multifragmentation mechanism is the
distribution of fragment sizes (Z), of relevance to the question of
critical phenomena and the liquid-gas phase transition.
The ISiS charge distributions have been analyzed in terms of a
power-law function, $\sigma(Z) \propto Z^{-\tau}$, shown in
Fig. \ref{Fig21}.  The results are nearly identical for all of the
hadron-induced reactions and behave similarly for the $^3$He +
$^{197}$Au data \cite{Fo96}.  At the lowest excitation energies, the
large values of the power-law exponent $\tau$ imply that small
fragments dominate the charge distribution, consistent with
lower-energy proton-induced reactions \cite{Gr84,Ye90,Er91}.  As the
system is heated, $\tau$ values decrease, signifying the increasing
tendency to form larger clusters.  A minimum in $\tau$ is reached near
E*/A $\approx$ 5-6 MeV, corresponding to the rapid increase in IMF
multiplicities. The tendency to form lighter clusters at high
excitation energy is most likely due to the dissolution of the larger
clusters in the heat bath and/or the formation of highly-excited
clusters that undergo secondary decay.  The average
variances of the Z distributions have also been measured \cite{Br98}
and the relation of both $\tau$ and the average variance is discussed
in Sec.6 with regard to phase transition arguments.

\subsubsection{Cross Sections}

Comparison of the excitation-energy distribution in Figs.\ref{Fig7}
and \ref{Fig14} and the IMF multiplicity probabilities in
Fig. \ref{Fig20} demonstrates that multifragmentation is a small
fraction ($\leq$ 10\%) of the total cross section in these light-ion
reactions, most likely originating in the lowest impact-parameter
collisions.  For the $^3$He beams the calculated total reaction cross
section is approximately 2000 mb for the $^{197}$Au target and 1400 mb
for $^{nat}$Ag \cite{Ka75,Kox84}. For the hadron-induced reactions,
the reaction cross section is about 1900 mb \cite{Ka75,Kox84}.  The
multifragmentation (M$_{IMF} \geq$ 3) cross section ranges from about
50 mb for the 5 GeV/c $\pi^-$ beam to 100-125 mb at the higher
energies, where excitation-energy saturation occurs.

In Table 2 the cross section dependence on target, beam energy and IMF
multiplicity is presented for the $^3$He-induced reactions.  The
yields clearly increase with target mass and projectile energy, with
cross sections that range from 9 mb for the 1.8 GeV $^3$He +
$^{nat}$Ag case to 190 mb for the $^{197}$Au target at 4.8 GeV.  The
effect of excitation-energy saturation in the $^{nat}$Ag system near
3.5 GeV bombarding energy is evident.

\begin{table}
\caption{Cross sections for the $^3$He-induced reactions}
\label{table2}
\begin{tabular}{ccccccc}
 & \multicolumn{2}{c}{\underbar{$^3$He + $^{nat}$Ag}} 
& \multicolumn{3}{c}{\underbar{$^3$He + $^{197}$Au}} \\
E$_{beam}$ (GeV) & 1.8 & 3.6 & 4.8 & 1.8 & &  4.8 \\
M$_{IMF}$  & & & cross section (mb) & & & \\
1 & 140 & 160 & 190 & 270 & & 300 \\
2 & 28 & 77 & 98 & 66 & & 170 \\
3 & 7.4 & 26 & 28 & 14 & & 110 \\
4 & 1.3 & 6.4 & 6.5 & 2.9 & & 54 \\
5 & 0.2 & 1.2 & 1.2 & 0.5 & & 20 \\
6 & 0.03 & 0.2 & 0.2 & 0.1 & & 6.6 \\
7 &  -    & -   &  -  &  -  & & 1.8 \\
8 &  -    & -   &  -  &  -  & & 0.4 \\
9 &  -    & -   &  -  &  -  &  & 0.1 \\
10 &  -    & -   &  -  &  -  & & 0.02 \\
   --   & --   & --  & --  & -- & \\
$\sigma(M_{IMF}\geq 3)$ & 8.9 & 34 & 36 & 18 & & 190 \\
$\sigma_{IMF}(M\geq 1)$ & 170 & 270 & 320 & 350 & & 660 \\
$\sigma_{IMF}$(total) & 220 & 430 & 490 & 460 & & 1300
\end{tabular}
\end{table}

\subsubsection{Source Charge}

\begin{figure}
\vspace{25mm}
\centerline{\psfig{file=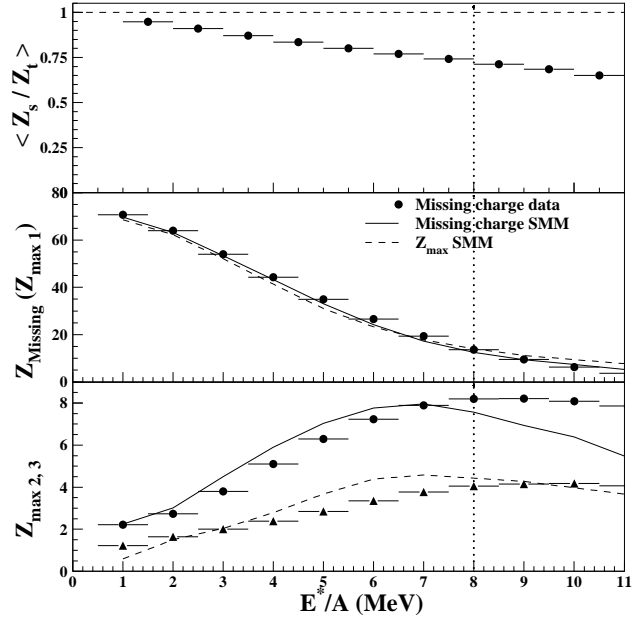,width=3.5in}}
\caption{Dependence of fractional source charge and IMF charges as a
function of E*/A for the 8 GeV/c $\pi^-$ + $^{197}$Au reaction.  Top:
fractional source charge.  Middle: missing charge in ISiS,
assumed to be the largest fragment, and SMM prediction for missing
charge (solid line) and for largest fragment (dashed line), both
passed through the ISiS filter. Bottom: charge of two largest observed
fragments, solid line is the SMM prediction for second largest fragment
(Z$_{max2}$) and dashed line for third largest fragment (Z$_{max3}$).}
\label{Fig22}
\end{figure}

Determination of the Z and A of the emitting source, as well as the
unmeasured heavy residue(s), is also an important component of the
reconstruction process described in Sec. 4.1.  In the top frame of
Fig. \ref{Fig22} the average fraction of the source charge relative to
the target charge is shown as a function of E*/A. As the excitation
energy increases, the effect of nonthermal particle emission becomes
quite strong, leading to average source charges of Z $\sim$ 60 at E*/A
$\sim$ 8 MeV.  The middle frame of Fig. \ref{Fig22} shows the average
missing charge in our reconstruction procedure, assumed here to be a
single fragment.  Comparison with the SMM model \cite{Bo90} gives good
agreement and these values closely correspond to the largest fragment
distribution observed in the EOS 1 GeV $^{197}$Au + $^{12}$C reaction
(when corrected for differences in thermal excitation
energy). \cite{Ha96}.

Beyond E*/A $\gtrsim$ 6 MeV the assumed fragment corresponding to the
missing charge (Z$_{max1}$) is an IMF (Z $\lesssim$ 20).  Finally, in
the bottom frame of Fig. \ref{Fig22} the average charges of the second
(Z$_{max2}$) and third (Z$_{max3}$) largest fragments are shown as a
function of excitation energy.  The sizes of the second and third
largest fragments remain nearly constant above E*/A $\sim$ 5-6 MeV, in
line with the results of Fig. \ref{Fig21}, and are also in relative
accord with SMM predictions up to E*/A $\sim$ 7-8 MeV.  At higher
excitation energies the data and SMM diverge, most likely due to the
storage of excess excitation energy in fragments in the model, leading
to secondary decay.  This divergence suggests that the fragments are
emitted relatively cold, as argued in \cite{Hu04}.

In summary, the thermal observables from the ISiS data present a
picture of a system that decays isotropically from a source with
velocity $\sim$ 0.01 c.  Above E*/A $\sim$ 5 MeV multifragmentation
(M$_{IMF} \geq$ 3) becomes the dominant decay mode and the spectra
suggest emission from a dilute/expanded source.  Near this energy the
largest fragments are formed.  Thus, these signals are qualitatively
consistent with expected observables from a liquid-gas phase
transition and in the following sections we examine this question in
greater depth.

\subsection{Breakup Density and Expansion}

A knowledge of the dependence of nuclear density on thermal excitation is of central
importance to our understanding of nuclear compressibility and the
equation-of-state of finite nuclear matter.  In addition, the breakup
density is particularly relevant to models of multifragmentation
phenomena, which assume that at sufficiently high temperatures,
thermal pressure and Coulomb forces drive nuclear expansion and
subsequent disintegration of the system
\cite{Bon85,Bo90,Du92,Fr90,Gr90}.  Perhaps the most direct experimental
signal of the breakup density is provided by the centroids of the
peaks of the IMF kinetic energy spectra as a function of
E*/A, as mentioned in previous sections and in
\cite{Po71,Po89,Fo96,Le01,Ye90,Wi92}.

\begin{figure}
\vspace{25mm}
\centerline{\psfig{file=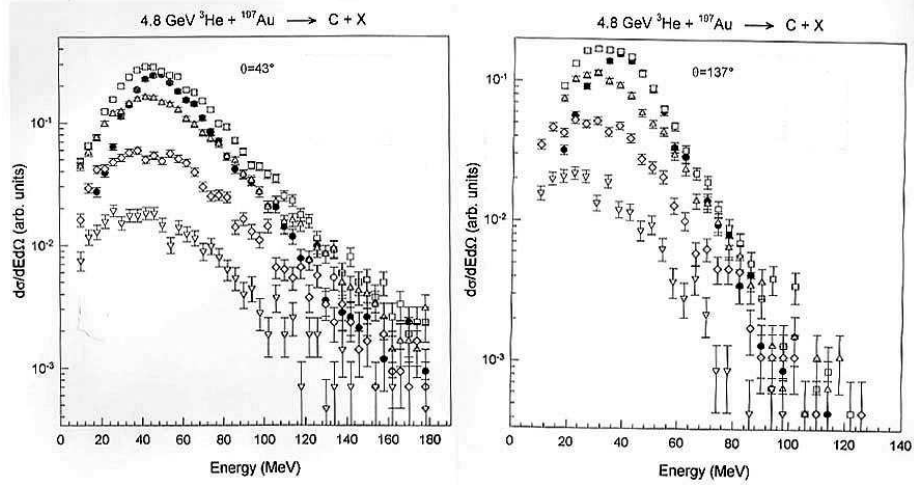,width=4.8in}}
\caption{Energy spectra at 43$^\circ$ and 137$^\circ$ for carbon
fragments emitted from the 4.8 GeV $^3$He + $^{197}$Au reaction as a
function of excitation energy.  Symbols are as follows:$<$E*/A$>$ = 3.4 MeV
($\bullet$); 4.6 MeV ($\Box$); 5.7 MeV ($\triangle$); 6.8 MeV ($\Diamond$);
7.9 MeV ($\bigtriangledown$).}
\label{Fig23}
\end{figure}

Breakup densities have been derived from the systematic Coulomb shifts
of the spectral peaks for a series of IMF data sets that span the
excitation-energy range E*/A = 0.9 to 7.9 MeV. Inclusive data were
analyzed for the 200-MeV $^4$He + $^{197}$Au \cite{Zh97} and E/A =
20-100 MeV $^{14}$N + $^{197}$Au \cite{Wi92} reactions.  Exclusive
data were based on the 4.8 GeV $^3$He + $^{197}$Au reaction
\cite{Br04}.  The spectra for each system were measured with low
kinetic-energy thresholds and covered nearly the entire 180 degree
angular range.  The threshold/angle criteria are essential for
obtaining stable moving-source fits, which require very good
definition of the low-energy component of the spectra.  Details of the
analysis are discussed in greater detail in \cite{Br04} and
\cite{Vi04}.  Representative IMF kinetic-energy spectra for carbon
fragments from the 4.8 GeV $^3$He + $^{197}$Au system are shown in
Fig.\ref{Fig23} as a function of E*/A.  The decrease in the peak
centroids with increasing excitation, opposite of expectations for a
thermal source at normal density, is evident.

\begin{figure}
\vspace{25mm}
\centerline{\psfig{file=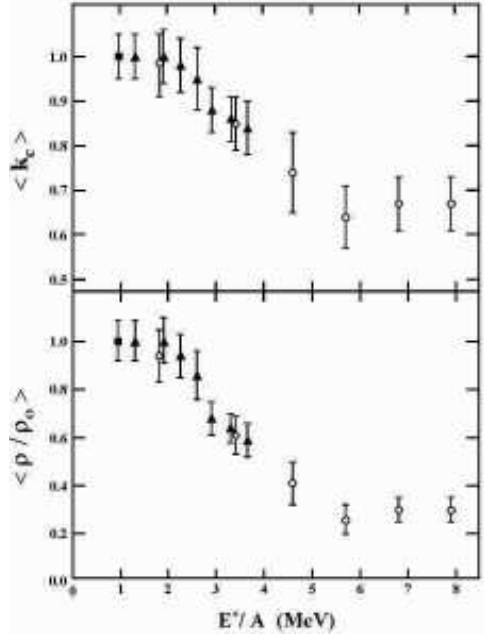,width=2.5in}}
\caption{Top: Dependence of the average fractional Coulomb
parameter $<k_C>$ as a function of excitation energy.  Symbols are as
follows: 200 MeV $^4$He (solid square);$^{14}$N (solid triangle); 4.8 GeV
$^3$He (open square).  Bottom: Average density $<$$\rho$/$\rho_0>$ 
as a function
of E*/A derived from the k$_{C}$ values in the top panel.}
\label{Fig24}
\end{figure}

The spectra were analyzed in terms of a two-component (three for
$^{14}$N) moving-source model \cite {We78} consisting of a
thermal-source described by a transition-state formalism
\cite{Mo75,Kw86}, a nonequilibrium source that assumes a Maxwellian
shape, and for $^{14}$N, a projectile-breakup source.  The thermal
source, of primary interest for this analysis, included the following
parameters: the source velocity, a fractional Coulomb term k$_C$, a
spectral slope temperature and a barrier fluctuation variable.  The
decreasing Z of the source with E*/A was also taken into account.
This formalism is designed primarily for binary breakups.  For
multifragmentation events, it is assumed that this procedure provides
a first-order approximation to the Coulomb field that exists between a
given IMF and the average of the residual nucleons.  For density
determinations the average fractional Coulomb parameter $<$k$_C$$>$
for IMFs is the sensitive parameter and is determined relative to
fission fragment kinetic energy systematics \cite{Br04,Vi04}. In the
top frame of Fig. \ref{Fig24} values of the Coulomb parameter are
plotted versus E*/A, where k$_C$ = 1 corresponds to nuclei at normal
density $\rho_0$.  Most striking about Fig. \ref{Fig24} is the sharp
decrease in $<k_C>$ in the excitation-energy interval E*/A $\approx$
2-4 MeV, suggesting emission from a source with a modified Coulomb
field.  It is also of note that this decrease matches the major
changes in other reaction variables, as shown in Sec. 4.2.

In order to obtain the breakup density, it is assumed that for these
light-ion-induced reactions the thermal source is spherical and
expansion is radially symmetric.  In this case, from Coulomb's law the
density expression reduces to

\begin{equation}
<\rho>/\rho_0 = k_C^3 ,
\end{equation}

\noindent since $<k_C>\alpha$$<r_C>^{-3}$, where r$_C$ is the mean
separation distance at breakup.  The result is shown in Fig. \ref{Fig24} where
$<$$\rho>$/$\rho_0$ is plotted as a function of E*/A.  Up to E*/A $\approx$
2 MeV the density appears to correspond to normal density.
Between E*/A $\approx$ 2 and 5 MeV, Fig. \ref{Fig24} indicates a
systematic decrease in density from $<$$\rho>$/$\rho_0 \approx$ 1.0 to
$<$$\rho>$/$\rho_0$ $\approx$ 0.3.  Above E*/A $\gtrsim$ 4 MeV, a nearly
constant value of $<$$\rho>$/$\rho_0\lesssim$ 0.30 is found within 
experimental error.

In summary, this spectral-shape analysis indicates that above E*/A
$\gtrsim$ 2 MeV, nuclear breakup occurs from an increasingly expanded/dilute
configuration.  Beyond E*/A $\gtrsim$ 4 MeV, a value of
$<$$\rho$/$\rho_0>\approx$ 0.3 is found, consistent with the predictions of
multifragmentation models \cite{Bo90,Fr90}.  Finally, the 
relative constancy of $<$$\rho$/$\rho_0>$ at high excitation 
energies suggests that a limiting breakup density has been 
reached \cite{Bn84,Na02}. 

The energy that drives expansion and subsequent multifragmentation is
usually attributed to either internal thermal pressure \cite{De93} or
the response to compressional forces produced in the early stages of
the target-projectile interaction \cite{Re97}.  As discussed in
Sec. 3, model calculations of the reaction dynamics for the systems
studied in this work provide little or no evidence for
compression-decompression effects.  Instead, the fast cascade creates
an initial residue with lower than normal nuclear density and high
thermal energy.  Thus, in highly asymmetric collisions at GeV
energies, only the thermal pressure and Coulomb field are of primary
relevance to the subsequent expansion process.

Two stages of thermal expansion leading to multifragmentation can be
schematically defined.  The first drives the nucleus to the breakup
configuration, where repulsive Coulomb forces exceed the restoring
nuclear force.  Although the breakup density beyond this point may
remain nearly constant, as in Fig. \ref{Fig24}, the increasing heat
content of the source may lead to an additional source of radial
expansion energy (or flow); i.e., the thermal energy that exceeds the
minimum necessary to reach the breakup density, defined here as excess
expansion energy, $\epsilon_{th}$. The impact of the excess expansion
energy on the spectra will be an increased flattening of the
high-energy spectral slope, over and beyond that expected for the
freezeout density and temperature \cite{Le00}.

In order to investigate the possible contribution of $\epsilon_{th}$,
it is necessary to choose a reference point that accounts for thermal
and Coulomb contributions to the fragment kinetic energies.  For this
purpose, three statistical models have been employed: SMM
\cite{Bo90,Bo95}, SIMON-explosion and SIMON-evaporation \cite{Ma97}.
In order to minimize any contamination from possible preequilibrium
emissions in the data, we compare the calculations with the measured
average fragment kinetic energy, for which preequilibrium
contributions to the tails of the spectra are negligible.  The inputs
to all three models were the same, using the source charge, mass,
velocity and excitation-energy distributions for the reconstructed
data (Sec.4.1), and then passed through the ISiS filter.

\begin{figure}
\vspace{30mm}
\centerline{\psfig{file=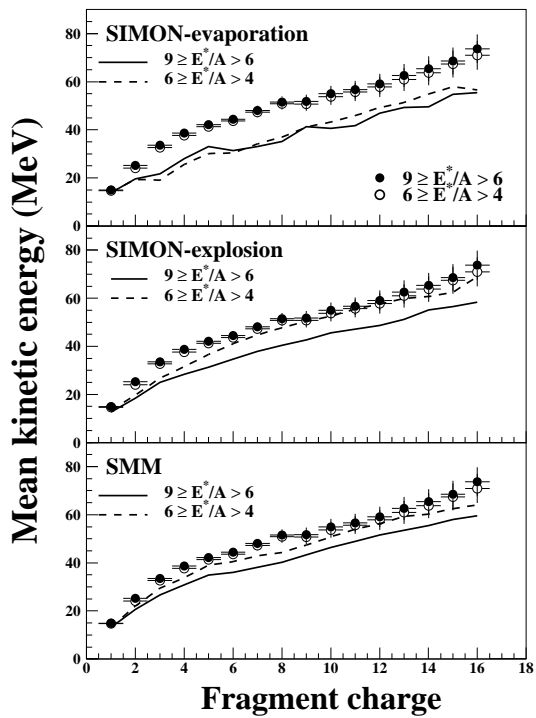,width=3.0in}}
\caption{Comparison between experimental and simulated fragment mean
kinetic energies calculated for two bins in excitation energy.  In
each panel, data are shown with open and solid circles and simulations
with dashed and solid lines.  The corresponding bins of excitation
energy are indicated on the figure.  SMM and SIMON-explosion
calculations have been performed without additional expansion energy.}
\label{Fig25}
\end{figure}

\begin{figure}
\vspace{25mm}
\centerline{\psfig{file=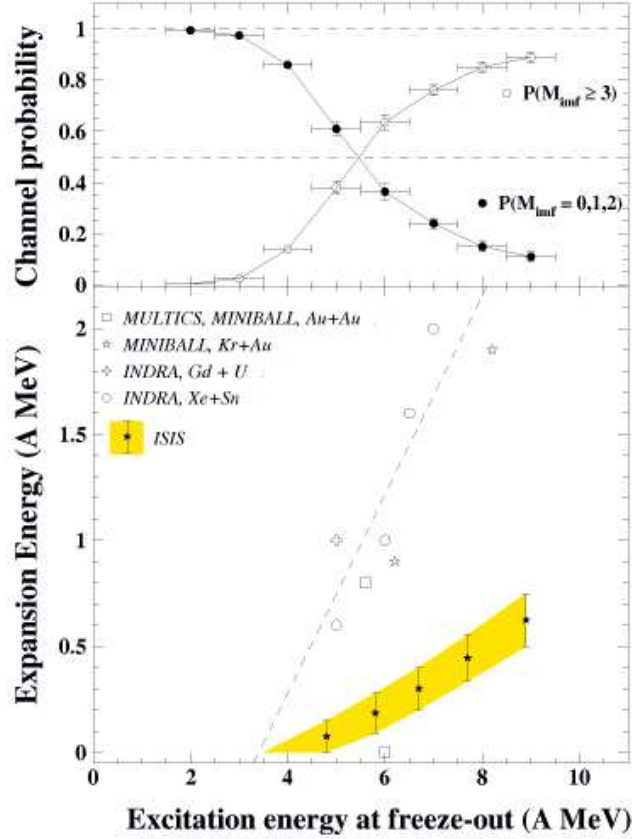,width=3.5in}}
\caption{Upper panel: See Fig. 20.  Lower panel: comparison 
between 8 GeV/c $\pi^- + ^{197}$Au reactions
and central heavy-ion collisions \cite{Le00}. 
The shaded area corresponds to the
ISiS excess expansion energies extracted with SMM at 3V$_0$ (upper limit)
and 2V$_0$ (lower limit).  The dashed line summarizes the excess expansion
energies extracted in central heavy-ion collisions with various
assumptions regarding the source characteristics.}
\label{Fig26}
\end{figure}

In Fig. \ref{Fig25} the calculations are compared with mean fragment
kinetic energies as a function of fragment charge for
excitation-energy bins, E*/A = 4-6 and 6-9 MeV.  The evaporative model
underpredicts the data substantially, although it does give reasonable
agreement for E*/A $\lesssim$ 3 MeV.  Both of the simultaneous
multifragmentation models describe the mean kinetic energies, as well
as the multiplicity and charge distributions, for E*/A = 4-6 MeV bin.
However, for the E*/A = 6-9 MeV bin both models fall below the data.
This shortfall is attributed to the existence of excess thermal
expansion energy.  Using the SMM model as a reference point, the
excess expansion energy is extracted from the difference between the
model and the data.  In Fig. \ref{Fig26} the results are plotted
versus E*/A.  This analysis indicates that the $\epsilon_{th}$
threshold occurs near E*/A $\cong$ 4 MeV and then gradually increases
to 0.5 A$_{IMF}$ MeV at E*/A = 9 MeV.  This amount of energy, while
small, must be taken into account when performing the calorimetry
(Sec. 2).  In contrast, central heavy-ion reactions show a much more
dramatic increase in the excess expansion energy, also shown in
Fig. \ref{Fig26}, suggesting that the excess expansion energy observed
in heavy-ion collisions may be related to the dynamical stage, perhaps
due to initial compression.

To summarize, moving-source analyses of the IMF spectra show a
systematic downward shift in the Coulomb peaks, supporting a picture
in which the breakup density decreases as a function of excitation
energy, even after correcting for nonequilibrium charge loss. While the density remains nearly constant above
E*/A = 5 MeV, the flattening of the spectral slopes suggests a small,
but measurable excess breakup energy that increases nearly linearly as $\epsilon_{th}$/A$_{IMF}$ = (0.1 E*/A)-0.4 MeV.

\subsection{Breakup Time Scale}

Central to any interpretation of multifragmentation events in terms of a
liquid-gas phase transition is the question of time scale.  For
evaporative cluster emission from the liquid phase at low excitation energies,
fragments are produced from the surface via a binary sequential decay
mechanism.  This process requires relatively long emission times of
order 1000 fm/c at low excitation energies \cite{Ch88}. 
In contrast, when the spinodal boundary of the phase
diagram is crossed, the system falls apart on a near-simultaneous time
scale via bulk emission from the entire nuclear volume.

Information about the emission time scale can be extracted by means of
the intensity-interferometry technique, which probes the mutual Coulomb
repulsion between fragment pairs emitted in proximity to one another in
space and time \cite{Boa90,Tr87,Ki91,Gl94}.  This technique
constructs the correlation function R for fragment pairs of reduced
velocity ${\rm v}_{red}$,

\begin{equation}
{R(\rm v_{red}) + 1 = C \frac{N_{corr}(\rm v_{red})}
{N_{uncorr}(\rm v_{red})}}.
\end{equation}

\noindent N$_{corr}$ is the
measured coincidence yield, while N$_{uncorr}$ is the uncorrelated
yield calculated with the event-mixing technique \cite{Ki91}, and the
normalization C is performed relative to the integral yields of each
\cite{Be00}.

\noindent The reduced velocity is given by

\begin{equation}
\rm v_{red} =\frac{\mid \rm v_1-\rm v_2 \mid}{\sqrt{Z_1 + Z_2}},
\end{equation}

\noindent where v$_i$ and Z$_i$ are the laboratory velocity and charge of the
fragments, respectively.  The denominator permits comparison of
different IMF Z values.  

\begin{figure}
\vspace{25mm}
\centerline{\psfig{file=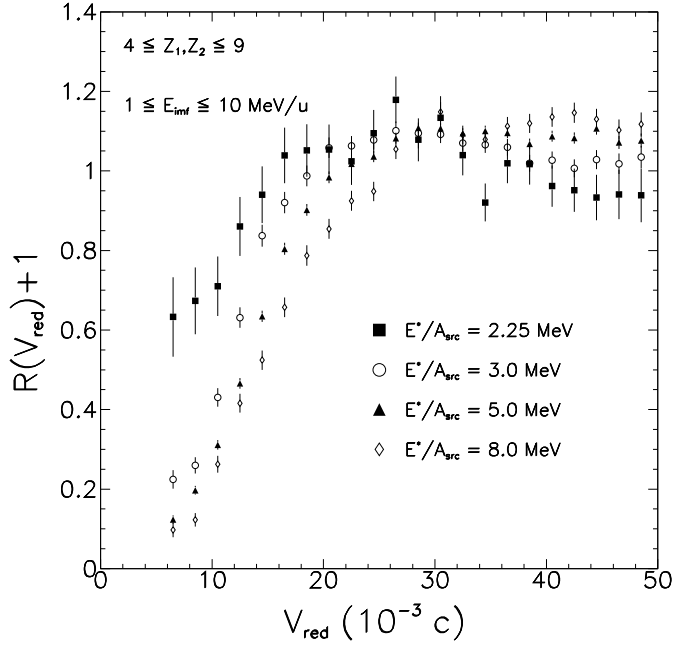,width=3.5in}}
\caption{Reduced-velocity correlation functions generated for four
different excitation energy energy per nucleon bins.  IMF kinetic
energy acceptance in the source frame is E$_{IMF}$/A = 1-10 MeV.}
\label{Fig27}
\end{figure}

Experimental IMF-IMF (4 $\leq$ Z $\leq$ 9) correlation functions from
hadron-induced reactions on $^{197}$Au between 8.0 and 11.2 GeV/c are shown
in Fig. \ref{Fig27} for several excitation-energy bins. Pairs emitted
in close proximity to one another in space and time (low v$_{red}$)
experience a supression in yield due to their mutual Coulomb
interaction (Coulomb hole).  Between E*/A = 2.25 and 5.0 MeV the
Coulomb hole increases, followed by a nearly constant supression at
higher excitation energies.  This effect is in qualitative agreement
with heavy-ion studies \cite{Ba93,Lo94}.

In order to extract the emission time scale, an N-body Coulomb
trajectory calculation \cite{Gl94,Pop98} has been performed, using the
experimental source and final product properties as input \cite
{Be00}.  The only adjustable parameters in the simulation are the
source volume, or separation distance between the residue and the
fragments.  The filtered output of the simulation must
reproduce both the small- and large-angle correlation data, as well as
the fragment charge distribution and kinetic energy spectra.  These
conditions impose a significant constraint on space-time ambiguities in
the simulation.  For purposes of calculating the
Coulomb energy, the separation distance is defined as

$$R_{12} = r_0 (A_{res}^{1/3} + A_{IMF}^{1/3}) + d$$

\noindent where r$_0$ = 1.22 fm.  Values of {\it d} between 2-6 fm
provide the best fits to the data and are consistent with the density
results described in the previous section.  The emission time {\it t}
is assigned via an exponential probability distribution, e$^{t/t_0}$,
where $t_0$ is the decay lifetime.

\begin{figure}
\vspace{25mm}
\centerline{\psfig{file=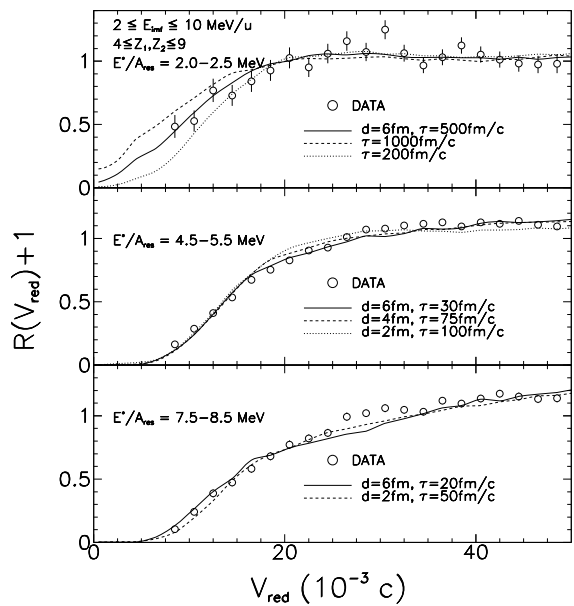,width=3.0in}}
\caption{Correlation function for Z = 4-9 IMFs as a function of
reduced velocity (open circles).  IMF kinetic energy acceptance in the
source frame is E$_{IMF}$/A = 2--10 MeV.  Data are gated on source E*/A =
2.0--2.5 MeV (top), 4.5--5.5 MeV (center) and 8.5--8.5 MeV
(bottom).  Solid and dashed lines are results of a Coulomb trajectory
calculation for fit parameters indicated on the figure.}
\label{Fig28}
\end{figure}

\begin{figure}
\vspace{25mm}
\centerline{\psfig{file=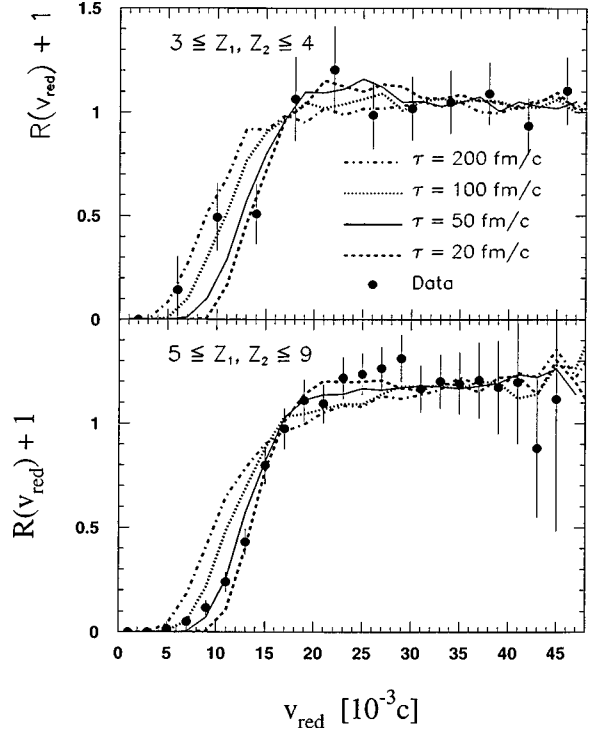,width=3.0in}}
\caption{Reduced-velocity correlations as a function of reduced
velocity for the 4.8 GeV $^3$He + $^{197}$Au reaction (points)\cite{Wa98}.  
Data were selected for pairs of events in which N$_{th} \geq$ 11 and
kinetic energy (E/A)$_{IMF}$ = 0.7-3.0 MeV and are shown for Z = 3,4 
fragments (upper frame) and Z = 5-9 fragments (lower frame).  
Lines are results of an
N-body simulation with $\rho /\rho_0$ = 0.25 and maximum residue size,
Z$_{res}$ = 12.  Time scales are indicated in the figure.}
\label{Fig29}
\end{figure}

In Fig. \ref{Fig28} the experimental correlation functions are
compared with simulations for a range of {\it d} and time values that
yield minimum chi-squared values.  Between E*/A= 2.0-2.5 and 4.5-5.5
MeV the emission time decreases by an order of magnitude, from $\sim$
500 fm/c to 20-50 fm/c.  Above E*/A $\sim$ 5 MeV the emission time
becomes very short and nearly independent of excitation energy,
consistent with a near-instantaneous breakup scenario.  Similar
results have been shown for the 4.8 GeV $^3$He + $^{197}$Au reaction,
as shown in Fig. \ref{Fig29} \cite{Wa98}.  Here the experimental
gating is performed for all fragments from events with E*/A $\geq$ 5
MeV.  Again the results are consistent with breakup times between
20-50 fm/c for that fraction of the yield that falls in the
multifragmentation regime.

\begin{figure}
\vspace{25mm}
\centerline{\psfig{file=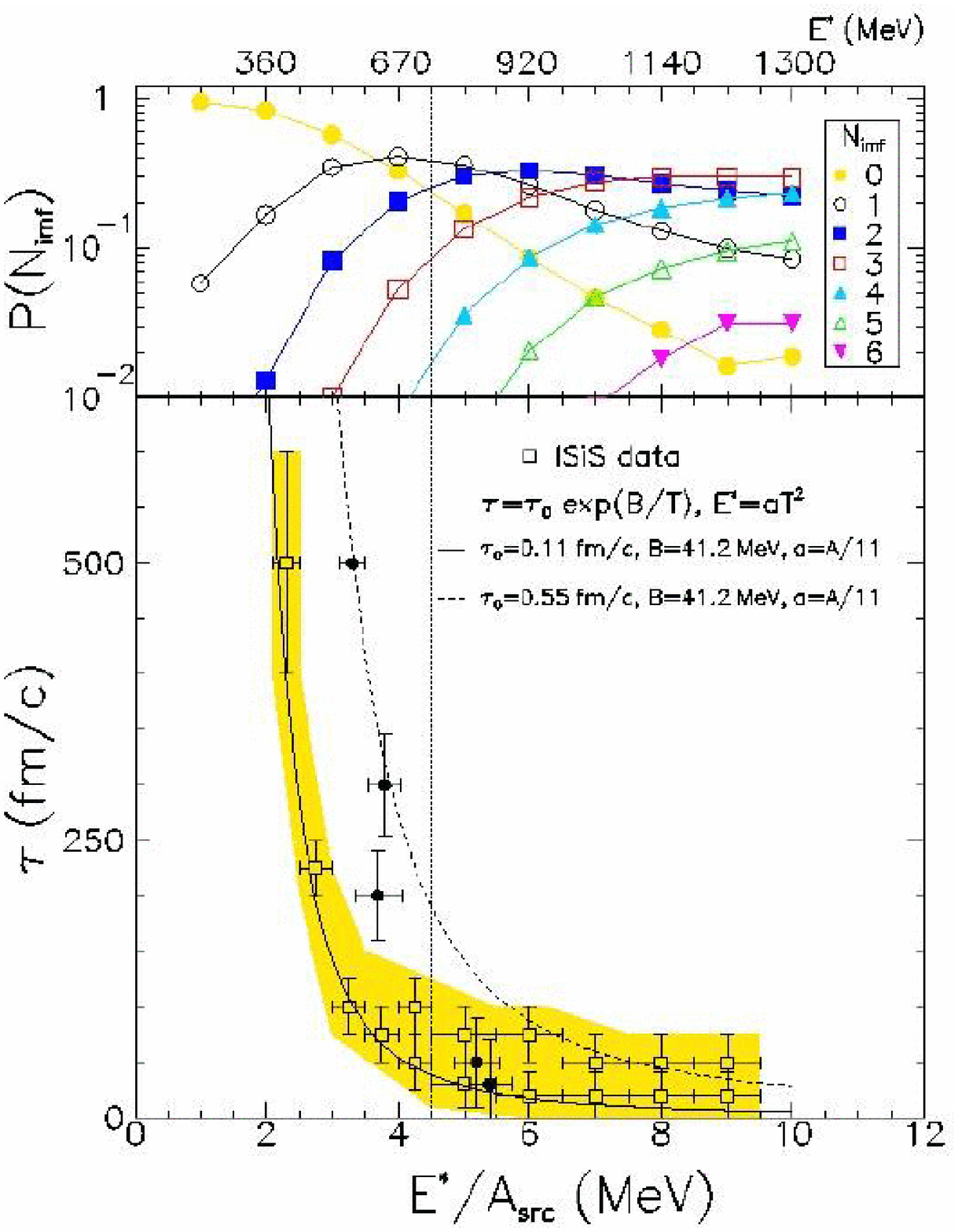,width=3.0in}}
\caption{Upper frame: IMF probability as a function of E*/A.  Bottom
frame: emission time as a function of E*/A.  Open points are ISiS data
under two extremes of fitting procedure.  Solid points are heavy-ion
data from \cite{Du98}.  Parameters for exponential fits to the data
are given on the figure.}
\label{Fig30}
\end{figure}

The lower panel of Fig. \ref{Fig30} presents the best-fit decay times 
{\it for events in which
two or more IMFs are emitted} in hadron-induced thermal
multifragmentation of $^{197}$Au nuclei. For reference the individual IMF multiplicities are shown in the upper panel.  The decay lifetimes at low
excitation energy are consistent with an evaporative mechanism, while
at higher energies the very short lifetimes support a
near-simultaneous breakup.  The shaded band in Fig. \ref{Fig30} covers
the range of space-time values that provide a consistent fit to all of
the observables.  Also shown in Fig. \ref{Fig30} are results for
heavy-ion reactions \cite{Du98}, which yield similar results, but
somewhat longer lifetimes at low energies.

In summary, the time scales derived from the intensity-interferometry
analysis demonstrates the evolution from the evaporative to
near-simultaneous breakup regime.  As with the multiplicity, spectra
and density evolution discussed in Secs. 4.1-4.3, the time scale
determinations provide a strong case for an interpretation in terms of
a transition from surface to bulk emission in the excitation energy
interval between E*/A $\sim$ 3-5 MeV.

\section{Thermodynamics}

\subsection{ The Caloric Curve:  Isotope-ratio Temperatures}

One of the most stimulating early results of multifragmentation
studies was the excitation energy versus temperature curve, or caloric curve,
proposed by the ALADIN group \cite{Po95}.  By plotting temperatures
derived from double-isotope ratios \cite{Al85} as a function of
excitation energy, a result was obtained that resembles the heating of
liquid water to the vaporization phase. Subsequent experiments,
including those described in this section, produced similar results
\cite{Na02}.  From the systematic behavior of these data, Natowitz has
derived a value of the critical temperature of 16 $\pm$ 1 MeV for
infinite nuclear matter and a nuclear compressibility constant K = 232
$\pm$ 30 MeV \cite{Na02a}.

In order to construct the heating curve for the ISiS data, the heat
content was based on the calorimetry described in Sec.4.1.  The
double-isotope-ratio technique for defining temperature is limited for
the ISiS data due to the high thresholds for isotope identification.
The isotope-ratio temperatures T were calculated according to the
prescription of Albergo \cite{Al85}, with correction factors $\kappa$
proposed by Tsang \cite{Ts97},

\begin{equation}
1/T = \frac{\rm{ln}(aR)-\rm{ln}\kappa}{B} .
\end{equation}

\noindent Here B is a binding-energy parameter, {\it a} is a statistical
factor dependent on ground-state spins, and R is the double-isotope
ratio.  For ISiS, useful isotope identification is restricted to LCPs
so that the relevant ratios are

\begin{equation}
R_{pd-He} = (p/d)/(^3He/^4He)\ \  \rm{and}
\end{equation}

\noindent

\begin{equation}
R_{dt-He} = (d/t)/(^3He/^4He),
\end{equation}

\noindent where all ratios involve only thermal LCPs.

\begin{figure}
\vspace{25mm}
\centerline{\psfig{file=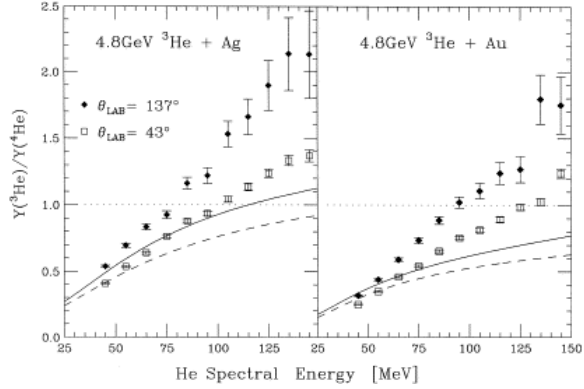,width=3.0in}}
\caption{He isotope ratios \cite{Kw98} as a function of He energy
observed at 43 and 137 degrees for the 4.8 GeV $^3$He reaction on Ag (left
panel) and Au (right panel). Lines are INC/EES model predictions
\cite{Fr90,Ya82} for 137 degrees (solid ) and 43 degrees (dashed).
Error bars are statistical only. }
\label{Fig31}
\end{figure}

The definition of thermal LCPs, as discussed in Sec. 4.1, can lead to
variability in the value of T that is obtained.  For p/d and d/t
ratios, there is only a small sensitivity to particle kinetic energy
\cite{Ru02,Br04}, so the distinction between thermal and
preequilibrium particles is of minor significance due to the
logarithmic nature of Eq. 5.1.  In contrast, as shown in
Fig. \ref{Fig31}, the $^3$He/$^4$He ratio increases strongly with kinetic
energy \cite{Kw98,Ru02,Br04}.  Thus the controlling factor in
determining T is the $^3$He/$^4$He ratio, which is dependent on the
thermal cutoffs assumed in the calorimetry.

\begin{figure}
\vspace{28mm}
\centerline{\psfig{file=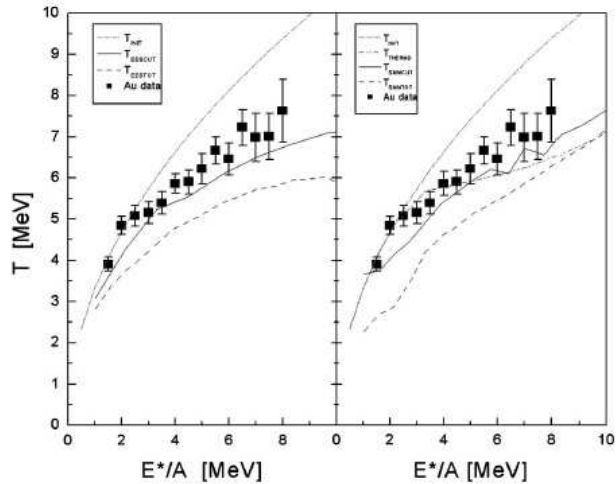,width=3.4in}}
\caption{(d/t)/($^3$He/$^4$He) isotope-ratio temperature
vs. reconstructed E*/A for the 4.8 GeV $^3$He + $^{197}$Au reaction.  Left
frame compares data with the INC/EES model \cite{Fr90,Ya79} and right
frame compares with the INC/SMM model \cite{Bo90,Ya79}.  Solid curves
are model predictions with experimental cuts imposed on H and He
kinetic energy spectra.  Dashed curves show the effect of removing the
experimental cuts.  Dotted curves show Fermi gas behavior with a=11
MeV$^{-1}$.  For the SMM case the dot-dashed curve gives the
thermodynamic temperature of the source.}
\label{Fig32}
\end{figure}

\begin{figure}
\vspace{30mm}
\centerline{\psfig{file=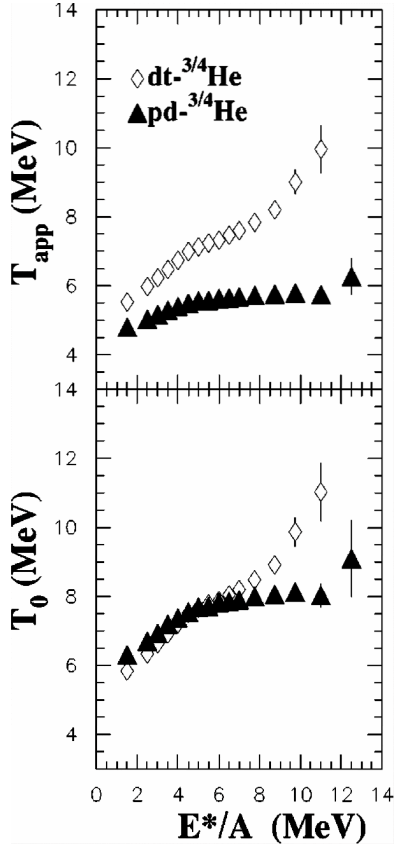,width=2.0in}}
\caption{Caloric curve for 8 GeV/c $\pi^-$ + $^{197}$Au from
(p/d)/ ($^3$He/$^4$He) and (d/t) $^3$He/$^4$He) thermometers 
using measured yields to calculate
temperature (top panel) and temperatures corrected for secondary decay
(bottom panel). }
\label{Fig33}
\end{figure}

In Figs. \ref{Fig32} and \ref{Fig33} the temperature versus E*/A
curves are shown for the 4.8 GeV $^3$He + $^{197}$Au system
\cite{Kw98} and the 8.0 GeV/c $\pi^-$ + $^{197}$Au system \cite{Ru02},
respectively.  For the $^3$He-induced reaction, which uses only the
dt-He ratio, there is no plateau, but a slope change is observed above
E*/A $\approx$ 2 MeV.  While the temperature increases from about T =
5-7 MeV in the E*/A $\approx$ 2-10 MeV range, it deviates markedly from a
simple Fermi gas prediction (dotted curves in Fig. 32).

Also shown in Fig.\ref{Fig32} are comparisons with INC/EES (Expanding,
Emitting Source) \cite{Ya79,Fr90} and INC/SMM \cite{Ya79,Bo90}
models. SMM comparisons assume the fragments are emitted cold;
comparisons with model parameters that produce hot fragments deviate
strongly from the data \cite{Ru02}.  For both model comparisons the
solid lines are predictions with the experimental cuts defined in
Sec.4.1 imposed on the model spectra.  The results provide fair
agreement with the data.  With the experimental cuts removed (dashed
curves), the caloric curves are lowered by $\sim$ 1-1.5 MeV per
nucleon, yielding better agreement with other caloric curves.  The
difference between the results with and without the experimental cuts
on the model is traced to the fact that the thresholds for isotope
identificaton in ISiS fall above the spectral peaks, where the
$^3$He/$^4$He ratio is much lower \cite{Wu79,Gr84} (i.e., R is larger
and T is smaller).  Also shown in the right frame of Fig. \ref{Fig32}
is the thermodynamic temperature predicted by the SMM model.

In Fig. \ref{Fig33} the caloric curves for the $\pi^-$ + $^{197}$Au
reaction are compared for the pd-He and dt-He thermometers.  The top
frame shows the results without the Tsang \cite{Ts97} corrections.
With the correction applied, both thermometers demonstrate a break in
the curve above E*/A $\approx$ 3-4 MeV, with the pd-He ratio yielding
a more distinct plateau.  At the highest excitation energies for the
dt-He case there is an indication of an upturn in the caloric curve
above E*/A $\approx$ 10 MeV, suggestive of possible entrance into the
vaporization regime.  Similar results have been observed in other
experiments and are also seen in the 4.8 GeV $^3$He data, although the
effect occurs at excitations where statistics are low and fluctuations
large.  However, this behavior is absent with the pd-He
thermometer. Thus, while the high E*/A dt-He results are intriguing, they are
not convincing.

The differences in temperature between the slope/plateau regions for the 4.8
GeV $^3$He and 8.0 GeV/c $\pi^-$ reactions can be traced to two
factors.  First, the kinetic-energy thresholds were lower in the
former case and second, the energy acceptance bins were not quite the same.
The net effect is that the $^3$He/$^4$He ratio is lower for the 4.8
GeV $^3$He measurements; i.e., R is larger and T lower.  

\begin{figure}
\vspace{25mm}
\centerline{\psfig{file=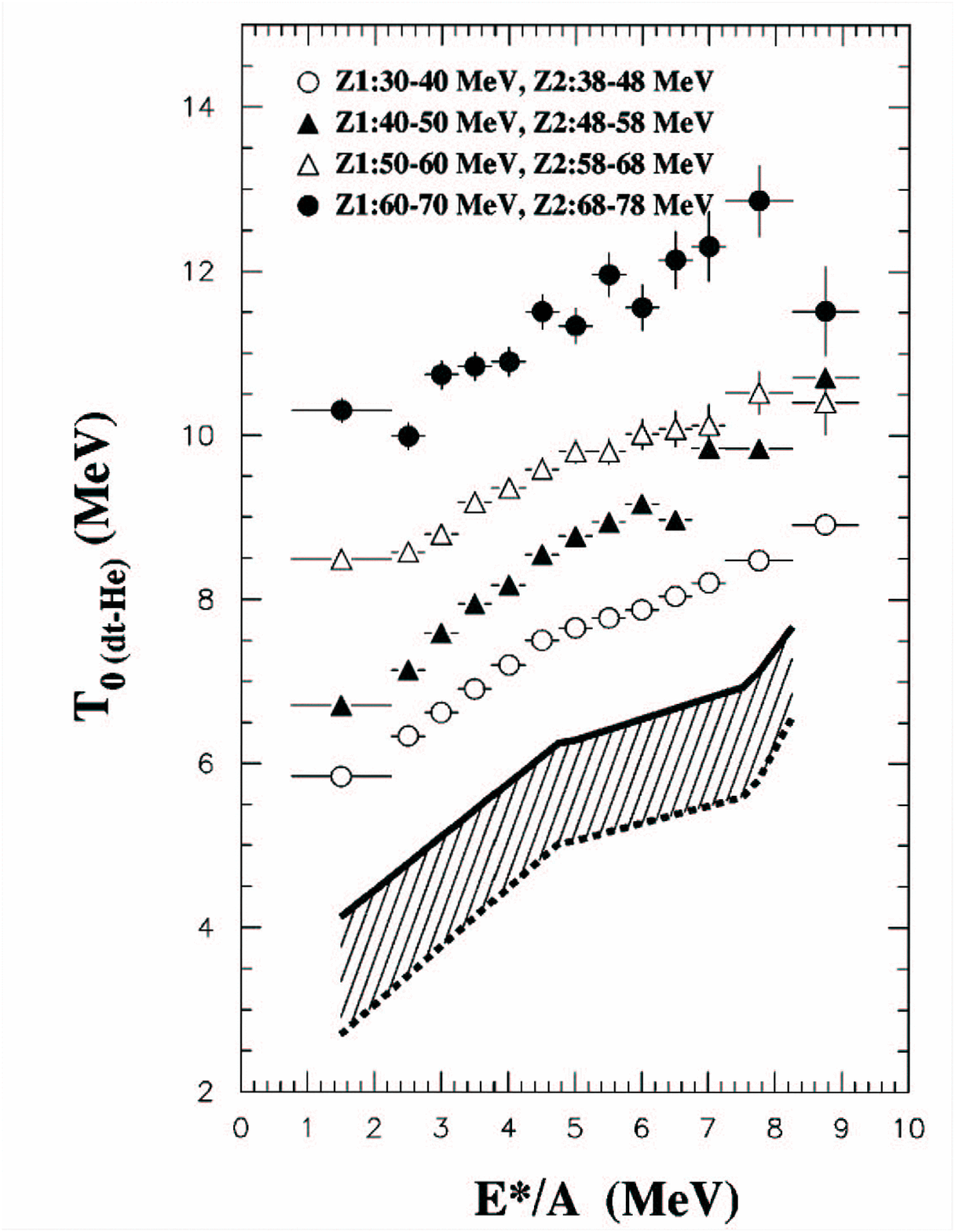,width=2.5in}}
\caption{The caloric curves for the 8 GeV/c $\pi^-$ + $^{197}$Au reaction 
from the
(d/t)/($^3$He/$^4$He) thermometer, corrected for secondary decay, using four
different kinetic energy acceptances as given in the graph.
The shaded area represents the caloric curve region extrapolated 
to the evaporative region of the H and He spectral peaks.}
\label{Fig34}
\end{figure}

The temperature dependences on $^3$He/$^4$He ratio can be used to
track the evolution of the de-excitation process leading up to
thermalization, under the assumption that the most energetic emissions
are emitted earliest \cite{Kw98-2,Ly00,Gu92}.  Such a ``cooling
curve'' is shown in Fig. \ref{Fig34} for the 8.0 GeV/c $\pi^-$ +
$^{197}$Au reaction, where Coulomb-corrected 10-MeV-wide bins have
been placed on LCP spectra \cite{Ru02}.  Note that the higher energy
bins correspond to the hard exponential tails of the spectra in
Fig. \ref{Fig12}.  The corresponding T vs. E*/A calculation reveals a
systematic decrease in the isotope-ratio temperatures as the kinetic
energy bin for the LCPs decreases.  The observed sequence of caloric
curves can be interpreted as evidence for the preequilibrium cooling
stage between the initial cascade and thermalization stages.  An alternative
explanation is provided by time-dependent EES model, for which
particles are emitted sequentially from an expanding, cooling source
\cite{Fr90}.

\begin{figure}
\vspace{25mm}
\centerline{\psfig{file=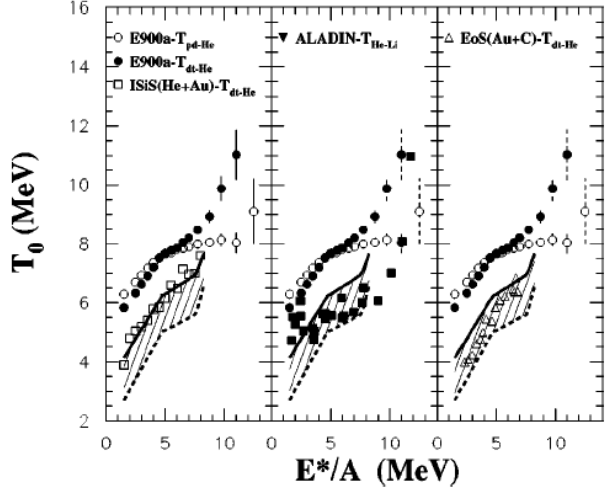,width=3.2in}}
\caption{Left:  Summary of caloric curves from the ISiS data in Figs.32-34, 
compared with ALADIN data \cite{Po95} (center) and EOS 
data \cite{EOS,Ha96} right.}
\label{Fig35}
\end{figure}

Due to the systematic trends of Fig. \ref{Fig34} and the lack of
mass-resolved data in ISiS in the Coulomb-peak region below
E/A = 8 MeV, the isotope- ratio temperatures would be lower if
extrapolated to the thermal LCP peak yields.  To examine this
correction, a linear fit was performed on the cooling curves of
Fig. \ref{Fig34} and then extrapolated to the Coulomb peak region for
element-identified LCPs, shown in Fig. \ref{Fig12}.  The shaded area
in Fig. \ref{Fig34} shows this extrapolation.  Figure \ref{Fig35}
compares the caloric curves from similar systems with the ISiS data.
The left-hand frame shows the difference between the observed and
corrected ISiS results and emphasizes the sensitivity of the
isotope-ratio thermometer to the energy acceptance for the LCPs. The
center- and right-hand frames compare the ISiS results with the ALADIN
peripheral Au + Au \cite{Po95} and EOS Au + $^{12}$C results \cite{EOS,Ha96} 
respectively.  Reasonable consistency is observed with the corrected
ISiS caloric curve and the systematics of Natowitz \cite{Na02}.

\subsection{The Caloric Curve:  Density-Dependent Fermi-Gas Temperatures}

The density determinations described in Sec. 4.3 suggest an
alternative approach to measuring the nuclear temperature.  Inherent
in the Fermi gas model is the first-order relationship

\begin{equation}
E^*/A = \frac{T^2}{K(\rho)}\bigl( \frac{\rho}{\rho_0} \bigr)^{2/3} 
\frac{m^*(\rho_o)}{m^*(\rho)},
\end{equation}

\noindent where K($\rho$) is the density-dependent inverse level
density parameter (1/a) and m* is the effective mass.  This predicted
dependence of temperature on density provides a method for determining
nuclear temperatures, {\it independent of isotope ratios}.  If one
assumes that the effective mass ratio is near unity at these high
excitation energies, then the ratio of K($\rho$) to K$_0$, the value
of K at normal density, becomes

\begin{equation}
K(\rho) = K_0 (\rho/\rho_0)^{2/3} = T^2/(E^*/A).
\end{equation}

\begin{figure}
\vspace{20mm}
\centerline{\psfig{file=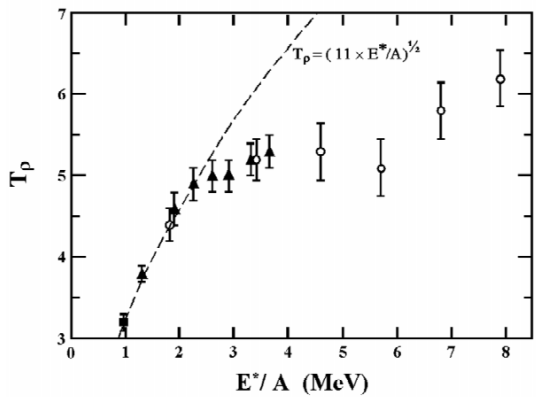,width=2.8in}}
\caption{First-order density-dependent Fermi-gas temperatures as a
function of excitation energy.  The dashed line is the normal-density
Fermi-gas prediction, with T = $\sqrt{11.3(E*/A)}$ MeV.  Symbols are
the same as in Figure 24.}
\label{Fig36}
\end{figure}

From fits to data below E*/A $\leq$ 2.0 MeV, an empirical
inverse level density parameter of $K_0$ = 11.3 MeV for a
density-independent Fermi gas is obtained, shown as a dashed curve in
Fig. \ref{Fig36}.  Using this value, Eq. 5.5 becomes

\begin{equation}
T(MeV) = [11.3 (\rho/\rho_0)^{2/3} (E^*/A)]^{1/2}.
\end{equation}

When the average densities derived from the IMF spectra [Sec. 4.3 and
Fig. \ref{Fig24}] are inserted into Eq. 5.6, the resultant
temperatures produce the caloric curve shown in Fig. \ref{Fig36}.  Up
to E*/A $\approx$ 2 MeV the temperature rises according to Fermi gas
predictions for nuclei at normal density.  In the region E*/A = 2-5
MeV a distinct slope change occurs, corresponding to the decrease in
breakup density of the emitting source.  Above E*/A $\approx$ 5 MeV,
Eq. 5.6 with a constant value of $\rho/\rho_0 \approx$ 0.30 predicts a
simple gradual increase in temperatures given by T = 2.2
$\sqrt{E^*/A}$.  Overall, the density-dependent Fermi gas model yields
a caloric curve that is strikingly similar to other caloric curve
measurements for similar reactions \cite{Po95,Ha96}, as well as the
corrected caloric curves from ISiS, shown in Fig. \ref{Fig34}.  This
result is consistent with statistical model calculations that assume
$\rho/\rho_0 \approx$ 1/3 at breakup \cite{Bon85,Bo90,Gr90} and in
qualitative agreement with the metastable mononucleus model of Sobotka
\cite{So04}.  A second-order analysis \cite{Vi04} of the density data,
in which expansion energy is taken into account, shows a 1-2 MeV dip
in the plateau near E*/A $\approx$ 5 MeV, but is otherwise similar.
Overall, the temperatures derived from the density-dependent Fermi gas
model and those from double isotope ratios \cite{Ts97} are in
agreement, lending additional support to the concept of caloric curve
behavior for hot nuclei.

\subsection{Heat Capacity}

\begin{figure}
\vspace{25mm}
\centerline{\psfig{file=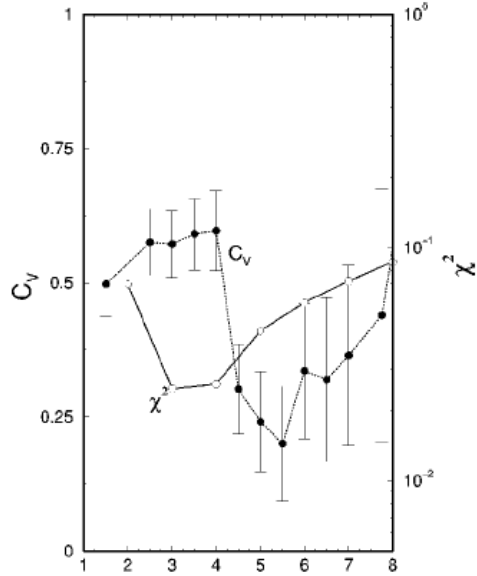,width=2.5in}}
\caption{Panel shows $C_{\rm v}$ and $\chi^2$
calculated from data.  The minimum of $\chi^2$
and the maximum of $C_{\rm v}$ coincide within experimental uncertainty.}
\label{Fig37}
\end{figure}

In heavy-ion studies evidence for a negative excursion in the
heat-capacity versus excitation-energy curve has been presented by the
MULTICS-MINIBALL Collaboration \cite {Ag00}.  Based on thermodynamic
considerations, this result provides possible evidence for a
first-order phase transition \cite{Gu99}.  The ISiS data have been
examined for such an effect \cite{Da02} and the results are shown in
Fig. \ref{Fig37}.  A sharp negative deviation in the heat capacity
C$_{\rm v}$ is observed near E*/A $\approx$ 4-5 MeV, consistent with the
heavy-ion results \cite{Ag00}.  Within experimental uncertainties the
minimum in the $\chi$-squared distribution coincides with the maximum
in the C$_{\rm v}$ curve.

The first-order phase transition argument is reflected in the
observations in earlier sections that show near E*/A $\approx$ 4-5 MeV
there is a sharp increase in the IMF multiplicity, a rapidly
decreasing emission time and density, the onset of excess expansion
energy, and a distinct slope change in the caloric curve.  The order
of the phase transition is discussed further in the following section.

In summary, both the corrected isotope-ratio and density-dependent
Fermi-gas temperature versus excitation energy plots show
quasi-caloric curve behavior with a transition in the region E*/A
$\approx$ 3-5 MeV.  Rather than a plateau, the resultant caloric
curves show a gradual increase in temperature with added heat, but are
well below normal Fermi-gas expectations.  By gating on LCP kinetic
energies, it is also possible to construct a ''cooling curve'' that
describes the evolution of the reaction mechanism from the cascade
step to thermalization.  In addition, a signal for a negative excursion
in the heat capacity curve supports heavy-ion data \cite{Gu99}
that are interpreted in terms of a first-order phase transition.

\section{The Liquid-Gas Phase Transition:  Scaling Law Behavior}

Given the overall agreement of the ISiS data with the qualitative
expectations for a liquid-gas phase transition, several further
questions of a more quantitative nature arise.  For example, do the
cluster size distributions behave according to statistical
expectations?  How well do the data conform to scaling laws for a
phase transition?  If so, what are the critical parameters and what is
the order of the phase transition?

Statistical behavior is an important question, not a priori obvious
for systems that evolve as rapidly as those formed in GeV
hadron-induced reactions.  At lower energies, where statistical
concepts are more appropriate, cluster emission probability
distributions can be well-described in terms of a binomial
distribution \cite{Mo95},

\begin{equation}
P^n_m (E^*)= \frac{m!}{n!(m-n)!} P^n(1-p)^{m-n}.
\end{equation}

\noindent Here,{\it n} is the IMF multiplicity, m is the number of chances
to emit an IMF, and {\it p} is the binary elementary probability.  The values
of {\it p} and {\it m} can be extracted from the experimental average
multiplicity $<N>$ and its variance,

\begin{equation}
<N> = mp \ \ {\rm and}\ \  \sigma^2_n = <N>(1-p).
\end{equation}

\begin{figure}
\vspace{25mm}
\centerline{\psfig{file=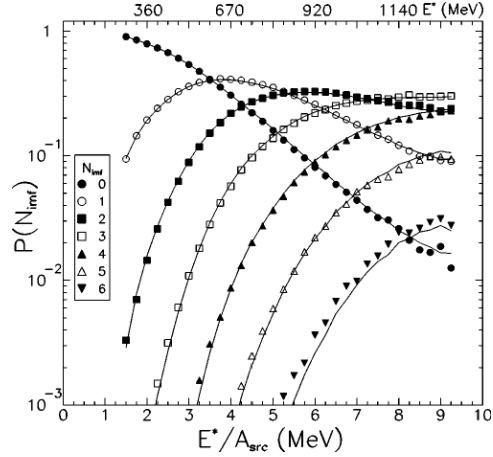,width=2.5in}}
\caption{The experimental (symbols) and calculated (lines) $n$-fold
IMF probability distributions as a function of E*/A.  The lines assume
a binomial probability distribution according to Eq. 6.1 and $p$ and
$m$ were extracted from Eq. 6.2.}
\label{Fig38}
\end{figure}

\noindent At lower energies it has been found that {\it p} is a
function of excitation energy, giving rise to the concept of thermal
scaling \cite{Mo95}. In order to test whether the ISiS results follow
this same statistical pattern, the 8.0 GeV/c $\pi^- + ^{197}$Au data
have been fit with Eqs. 6.1 and 6.2 \cite{Be01}.  The binomial distribution
analysis gives very good agreement with the data up to $N_{IMF}$ = 5,
as shown in the multiplicity distributions in Fig. \ref{Fig38}. Some
deviations appear for $N_{IMF} \geq$ 6, where statistics become
increasingly poor.  To investigate the possible dependence of the
binomial parameters on collision violence, the reciprocal of the
probability factor $p$ was plotted versus the total transverse energy,
thermal transverse energy and E*/A.  The total transverse energy $E_t$
diverges strongly at high energies, reflecting the contribution of
preequilibrium processes to the yield.  Removal of the preequilibrium
component yields an improved scaling fit to the data for the thermal
transverse energy. The strongest correlation is found when scaled as
a function of E*/A, where a nearly linear dependence is observed,
supporting an interpretation in terms of thermal scaling \cite{Be01}.

The evolution of the parameter $m$ with excitation energy is shown in
Table 3, where correlations with the size of the source Z$_{src}$,
total observed charge Z$_{obs}$ and the removal energy(-Q).  The one
variable that tracks most closely with $m$ is the removal energy,
suggesting that $m$ represents an energy constraint that allows only
certain partitions.

\begin{table}
\caption{Values of the binomial parameter $m$, the primary source size
A$_{src}$,
the observed charge Z$_{obs}$, and the Q$_{value}$ for
various E*/A bins.}
\label{table3}
\begin{tabular}{lcccccccc}
\hline\\
E*/A (MeV) & 2.0 & 3.0 & 4.0 & 5.0 & 6.0 & 7.0 & 8.0 & 9.0 \\
m & 3.36 & 4.82 & 5.61 & 6.29 & 6.72 & 7.67 & 8.31 & 7.79 \\
Z$_{src}$ & 74.3 & 71.2 & 68.4 & 65.8 & 63.5 & 61.1 & 59.0 & 57.1 \\
A$_{obs}$ & 8.75 & 15.2 & 21.9 & 28.1 & 33.5 & 38.1 & 42.2 & 45.4 \\
Q$_{value}$(MeV) & -160 & -204 & -249 & -295 & -341 & -383 & -420 & -451 \\
\hline 
\end{tabular}
\end{table}

The agreement with the thermal scaling concept indicates that {\it p}
can be expressed by a partial decay width

\begin{equation}
p = \frac{\Gamma}{\hbar \omega_o} = e^{-B/T}
\end{equation}

\noindent where $\omega_o$ is interpreted as the frequency of assault on
the barrier B at temperatures T \cite{Be01,Mo95}. Defining the
intrinsic emission time as $t_0$ = 1/$\omega_0$, the emission time is
given by

\begin{equation}
t = t_0 e^{B/T}, \rm{or}\ \   p = t_0/t .
\end{equation}

\begin{figure}
\vspace{25mm}
\centerline{\psfig{file=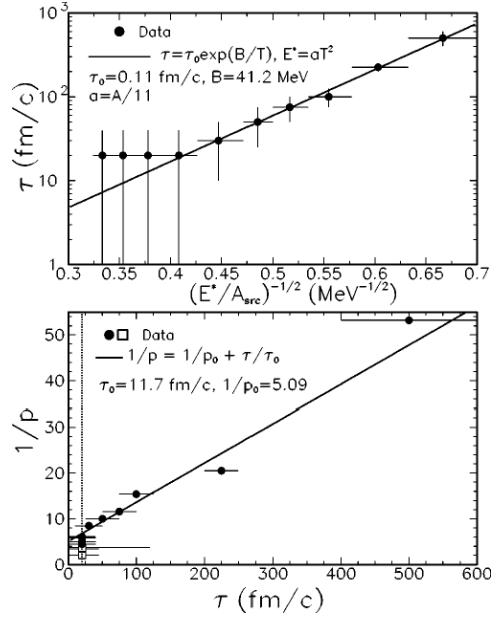,width=2.5in}}
\caption{Top panel: IMF emission time {\it t} as a function of (E*/A)$^{-1/2}$,
from Fig. 30.  The line corresponds to a fit using Eq. 6.4.  Bottom
panel: plot of 1/p vs time.  The solid line is a linear fit to the data. 
The dotted
line indicates the ``apparent'' saturation in emission time.}
\label{Fig39}
\end{figure}

\noindent Using the emission times derived in Sec. 4.4, a plot as a
function of the thermal scaling variable (E*/A)$^{1/2}$ shows nearly
linear behavior, as seen in the top frame of Fig. \ref{Fig39}. The
bottom frame of Fig. \ref{Fig39} shows the relation between the
lifetime {\it t} and 1/p.  A simple linear relationship is observed
down to emission times of 20 fm/c near E*/A $\approx$ 6 MeV.  The
evolution of the inverse probability 1/p at higher energies appears to
be independent of time, indicating a mechanism charge that favors a
space-like (bulk) emission scenario, rather than one that is
sequential, as is the case at lower excitations.

The apparent statistical nature of the thermal component of the data
justifies further investigation of expected liquid-gas phase
transition properties.  One method for extracting information relevant
to this issue is through a moment analysis of the fragment charge
distributions \cite{Fi82,Ca86}. Calculations with both percolation and
statistical multifragmentation models predict that the relative
variance $\gamma_2$ of the charge distributions will exhibit maxima of
$\gamma_2 \approx$ 2.1-2.3 near the critical point.  Brzychczyk
\cite{Br98} analyzed the relative moments of the 4.8 GeV $^3$He +
$^{197}$Au reaction and found a variance of $\gamma_2$ = 2.3 $\pm$ 0.1
near E*/A $\approx$ 5.5 MeV.  Thus, the ISiS data are in good
agreement with phase transition models and provide further 
consistency with arguments for a liquid-gas phase transition and 
possible critical behavior in hot nuclei.

Berkenbusch {\it et al.} \cite{Ber02} carried out a global percolation
analysis on the 10.2 GeV/c p + $^{197}$Au data from ISiS.  The
bond-breaking probability for the model is determined from the
excitation energy via the relation \cite{Ba85}

\begin{equation}
E_b (E^*) = 1 -\sqrt{\frac{2}{\pi}} \Gamma (3/2, 0, B/T(E^*))
\end{equation}

\noindent where $\Gamma$ is the generalized incomplete gamma function,
B is the binding energy per nucleon of the source and T is the source
temperature, determined from the excitation energy T = $\sqrt{E*/a}$
and a = A/13 MeV$^{-1}$.  Input to the model utilized experimental
values for the excitation energy and source size, with the lattice
size fixed by the size of the thermal residue.  An important feature
of the calculation is that account is taken of the secondary decay of
the excited primary fragments, which is particularly important for the
fragile Z = 3-5 fragments that comprise most of the IMF cross section.

\begin{figure}
\vspace{28mm}
\centerline{\psfig{file=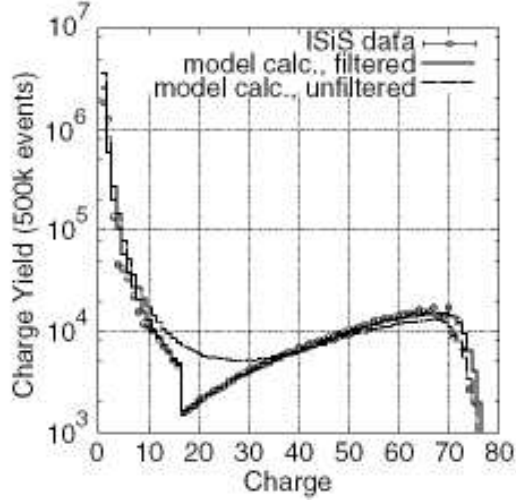,width=2.9in}}
\caption{Inclusive charge yield spectra for the reaction p +$^{197}$ Au at
10.2 GeV.  The round symbols represent the ISiS data.  The dotted
histogram is the result of the corresponding percolation model
calculation.  The thick histogram represents the output of the
calculations, filtered through the detector acceptance.}
\label{Fig40}
\end{figure}

Figure \ref{Fig40} compares the charge distributions from the data
with both filtered and unfiltered percolation values.  The
discontinuity just above Z = 16 is a consequence of the discrete
charge-identification limit in ISiS and the assumption that all
unmeasured charge resides in a single fragment. Overall, the filtered
percolation yields are in excellent agreement with the data.

For values of the control parameter {\it p} near the critical value, p$_c$,
the cluster number $n_s$ is predicted to scale as

\begin{equation}
n_s(p) = s^{-\tau} f[(p-p_c)S^\sigma]\ \ \rm{({\rm for}\  p \approx p_c)},
\end{equation}

\noindent where s is the size of the cluster (Z) and $\tau$ and
$\sigma$ are the two critical exponents of percolation theory.  The
scaling function {\it f} has the property that $f(0) = 1$; i.e., the
power-law dependence is valid only near $p = p_c$.  By associating
the bond-breaking probability with the temperature and assuming an
exponential function for {\it f}, the fractional IMF yield $<n_z>$
becomes

\begin{equation}
<n_z> = q_0Z^{-\tau} exp\bigl[\frac{(T-T_c)Z^\sigma}{T}\bigr])
\end{equation}

\noindent where $q_0$ is a normalization parameter and T$_c$ is the
critical temperature.  Thus, one expects a plot of

$$<n_z>/q_0 Z^{-\tau}$$

to scale exponentially with $(T-T_c)Z^\sigma/T$ for all fragment sizes.

\begin{figure}
\vspace{25mm}
\centerline{\psfig{file=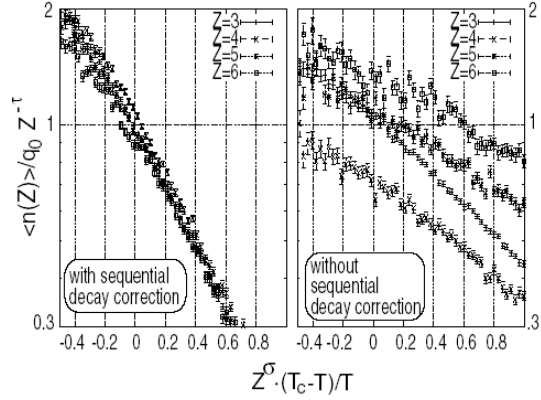,width=2.9in}}
\caption{Scaled fragment yields as a function of the scaled control
parameter for Z = 3, 4, 5 and 6.  The left-hand side shows the results
of inclusion of secondary decay corrections and the
right-hand side shows the fit when omitting these
corrections.}
\label{Fig41}
\end{figure}

The result of a $\chi^2$ optimization procedure for the theoretical
percolation charge distributions for this system yields values of the
critical parameters $\sigma = 0.5 \pm 0.1$ and $\tau = 2.18 \pm 0.01$,
in good agreement with percolation analysis for an infinite lattice,
$\sigma$ = 0.45 and $\tau$ = 2.18.  A similar analysis of the
Z$_{IMF}$ = 3-6 data from ISiS produces the results shown in
Fig. \ref{Fig41}.  The right frame shows the inability to achieve
scaling when sequential-decay corrections are omitted from the
analysis.  When corrected for sequential decay, as shown in the left
frame of Fig. \ref{Fig41}, much better scaling behavior is found.  The
critical exponents for this finite system, derived from the data, are
$\sigma = 0.5 \pm 0.1, \tau = 2.35 \pm 0.05$ and (E*/A)$_c$ = 5.3 MeV
(or T$_c$ = 8.3 MeV for a simple Fermi gas with $a$ = A/13
MeV$^{-1}$).

Since the infinite size limit of the model contains a continuous phase
transition for the range of excitation energies covered by the present
data set, the scaling agreement between the data and percolation
theory can be interpreted as evidence for a continuous phase
transition in nuclear matter.

The Fisher Droplet Model \cite{Fi67} provides another avenue for
examining the correspondence between IMF emission and a
liquid-gas phase transition.  Fisher's model describes the aggregation
of molecules in a vapor into clusters. The abundance of a given cluster size
A can be written as

\begin{equation}
n_A = q_0 A^{-\tau} exp \bigl[ (A\Delta\mu-Co \epsilon A^\sigma)/T\bigr],
\end{equation}

\noindent where in addition to the critical exponents $\tau$ and
$\sigma$ of Eq. 6.7, n$_A$ = $N_A/A_0$, the number of droplets of mass {\it A}
normalized to the system size $A_0$; the difference between the
actual and liquid chemical potentials is $\Delta_\mu = \mu-\mu_\ell$;
$C_o$ is the zero temperature surface energy coefficient, and
$\epsilon = (T_c - T)/T_c$.  This equation reduces to Eq. 6.7 if
$\Delta_\mu = 0$.

One approach taken by Elliott {\it et al.} \cite{El02} is to modify
Eq. 6.8 to take into account the Coulomb energy release when a
particle moves from the liquid to the vapor phase, which assumes

\begin{equation}
n_A = q_0 A^{-\tau} exp \bigl[ (A\Delta\mu +E_{Coul}-C_o 
\epsilon A^\sigma)/T\bigr],
\end{equation}

\noindent where $E_{Coul}$ is defined as follows

\begin{equation}
E_{Coul} = \frac{1.44(Z_{src} - Z_{IMF})Z_{IMF}}
{r_0\bigl[(A_{src}-A_{IMF})^{1/3} + A^{1/3}_{IMF}\bigr]} \times 
(1-e^{-\chi\epsilon}) \rm{fm-MeV}.
\end{equation}

\noindent Here $r_0$ = 1.22 fm and the exponent $\chi\epsilon$ insures
that the Coulomb energy disappears at the critical point.  Since IMF
masses are not measured in ISiS, it was assumed that A/Z = 2.
Temperatures were determined by the Fermi gas approximation of Raduta
\cite{Ra97}.  IMF acceptance was for IMFs with 5 $\leq Z \leq 15$, for
which preequilibrium effects are small \cite{El02}.  Subsequent 
modifications of the Fisher model
can be found in \cite{Mor-Pha05,Mo05}.

\begin{figure}
\vspace{25mm}
\centerline{\psfig{file=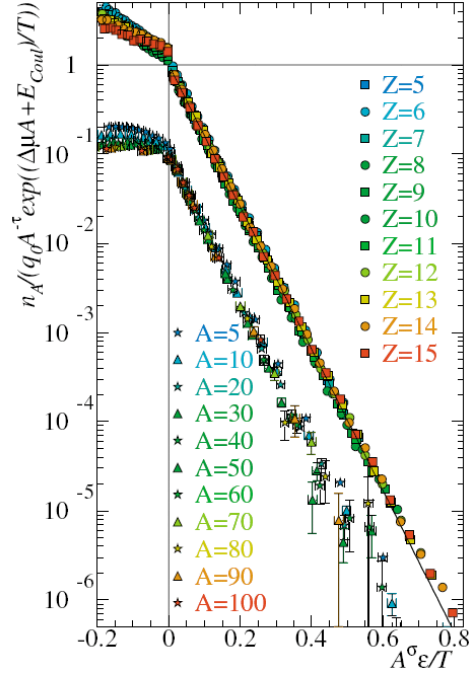,width=2.5in}}
\caption{The scaled yield distribution versus the scaled temperature
for the ISiS data (upper) and {\it d} = 3 Ising model calculation
(lower).  For the Ising model, the quantity $n_Z/q_0A^{-\tau}/10$ is
plotted against the quantity $A^{\sigma}\epsilon /1.435T$. 
Data for $T > T_c$ are scaled only as $n_A/q_0A^{-\tau}$.}
\label{Fig42}
\end{figure}

The results of scaling the data according to Eq. (6.9) are shown in
Fig.\ref{Fig42}, where the fragment mass yield is scaled by the
power-law prefactor, the chemical potential term and the Coulomb energy.
This quantity is scaled by temperature and the surface energy
parameter A$^\sigma_\epsilon$/T.  The scaled data follow Fisher
scaling over six orders of magnitude, which suggests that this line
represents the liquid-vapor coexistence line.  As supporting evidence,
also shown in Fig. \ref{Fig42} is the scaled cluster distribution from
a d=3 Ising model calculation \cite{Ma03} for a system that undergoes
a phase transition.

The values of the critical parameters from the $\chi$-squared
minimization routine are summarized in Table 4 for both the
percolation and Fisher scaling analyses.  Within errors, the Fisher
model values of $\tau$ = 2.28 $\pm$ 0.14 and $\sigma$ = 0.54 $\pm$ 0.01
are in good agreement with those from percolation as well as with
earlier work \cite{Da99,El00}. The surface-energy coefficient C$_o$ =
18.3 $\pm$ 0.5 MeV is in general accord with the liquid drop value of
16.8 MeV.  An important result relevant to previous scaling analyses
is that for the first time it has been possible to measure
$\Delta\mu$.  The measured value of $\Delta\mu$ = 0.06$\pm$ 0.03
substantiates the assumption that $\Delta\mu \cong$ 0 in previous
analyses. Similarly, the value of $x$ = 1.0 $\pm$ .06 insures that 
the Coulomb energy is small and does not affect the scaling significantly.

\begin{table}
\caption{Comparison of percolation and Fisher scaling resutls.}
\label{table4}
\begin{tabular}{ccc}
 \underline{parameter} & \underline{percolation} & \underline{Fisher} \\
 $\tau$ & 2.35 $\pm$ 0.05 & 2.18 $\pm$ 0.14 \\
$\sigma$ & 0.5 $\pm$ 0.1 & 0.54 $\pm$ 0.01 \\
T$_{crit}$ & 8.3 $\pm$ 0.2 MeV  & 6.7 $\pm$ 0.2 MeV \\
(E*/A)$_{crit}$ & 5.3 $\pm$ 0.3 MeV & 3.8 $\pm$ 0.3 MeV \\
C$_o$ & -- & 18.3 $\pm$ 0.5 MeV \\
$\Delta\mu$ & -- & 0.06 $\pm$ 0.03 MeV/A \\
$\chi$ & -- & 1.00 $\pm$ 0.06
\end{tabular}
\end{table}

Based on this analysis, the phase transition is first order up to the
critical point of excitation energy E$^*_c$/A =3.8 $\pm$ 0.3 MeV,
above which it becomes continuous.  Analysis of the EOS data \cite{EOS}
yielded a value of E$^*_c$/A = 4.75 MeV.  However, as discussed in
Sec. 3, when the excitation energies are calculated with the same assumptions for
elimination of preequilibrium particles, the ISiS and EOS critical
energies are the same.  The critical temperature for finite nuclei
derived from the Fisher scaling analysis is $T_c = 6.7 \pm$ 0.2 MeV.
Based on the Fisher scaling parameters derived from the data, it is
then possible to construct the two-phase (liquid-gas) coexistence line
over a large energy/temperature interval, extending up to the critical
point, from which the full phase diagram of nuclear matter can be
defined \cite{Fi67}.

In summary, the ISiS data are well-described by binomial reducibility
and thermal scaling analyses, providing a strong argument for the
statistical nature of multifragmentation.  Further, scaling with the
Fisher model can be used to define the liquid-gas coexistence line,
while a percolation analysis supports a continuous phase transition at
higher excitation energies.

\section{Summary and Conclusions}

Both the reaction dynamics and the subsequent decay of hot residues
formed in GeV light-ion-induced reactions on heavy nuclei have been
investigated with the ISiS detector array.  Of primary concern in this
effort has been the isolation and characterization of
multifragmentation events, believed to be the possible signature of a
nuclear liquid-gas phase transition in finite nuclei.

Bombardments with proton, antiproton, pion and $^3$He beams produce an
exponentially-decreasing distribution of excitation energies that
extend up to $\sim$ 2 GeV in reactions on $^{197}$Au nuclei.  The
deposition of excitation energy is found to increase as a function of
beam energy up to a momentum of about 8 GeV/c for reactions of hadrons
with $^{197}$Au and an energy of $\sim$ 4 GeV for the $^3$He + $^{nat}$Ag
system.  For higher beam energies there is little additional
increase in deposition energy, presumably due to a tradeoff between
beam energy and target transparency.  Relative to other hadron beams,
the optimum projectile for achieving high excitation energies is found
for 8 GeV/c antiprotons, for which the reabsorbtion of some fraction
of the decay pions can produce enhanced excitation of the residue.  For
the same beam momentum, proton and pion beams are nearly identical in
their excitation-energy distributions.

The fast LCP component of the spectra has been analyzed with a BUU
model that incorporates A = 2 and 3 nuclei in the scattering matrix.
Best fits to the data are obtained with a version that includes
modified in-medium scattering cross sections and a momentum-dependent
potential.  As shown in Figs. 4 and 5, the BUU calculations indicate
that for central collisions the hot residues are formed in a state of
depleted density -- due to fast knockout followed by preequilibrium
processes that occur on a time scale much faster that the relaxation
time.  Simulation of the time evolution of the collision dynamics
predicts that the entropy per nucleon becomes nearly constant after
about 30 fm/c, suggesting a randomized, but not necessarily
thermalized system.  From comparison of the BUU code with the fast LCP
spectra, it is inferred that a total time of about 60 fm/c is required to
reach a state of quasi-thermalization.

The most fundamental signature of a liquid-gas phase transition is the
observation of events in which an equilibrated hot nucleus
disintegrates into multple LCPs and IMFs.  For this purpose, thermal
events have been selected that are shown to decay isotropically in the
center-of-mass system with Maxwellian kinetic energy spectra from
which nonequilibrium components have been removed. 
The cross sections for these events is of order 100 mb.  

\begin{figure}
\vspace{25mm}
\centerline{\psfig{file=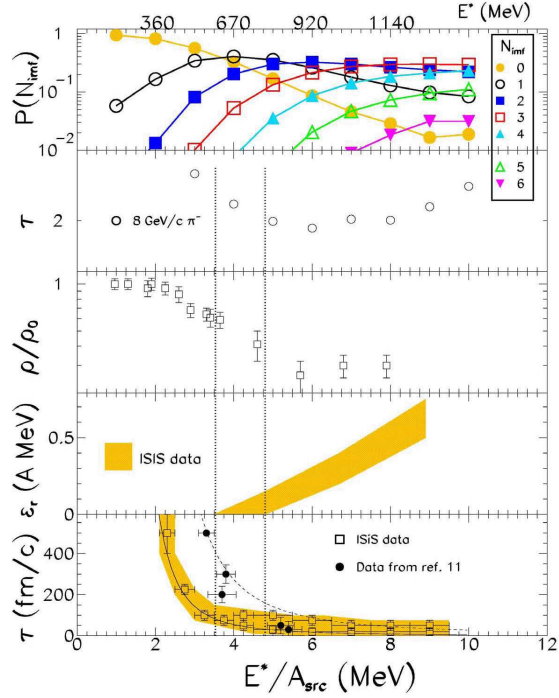,width=3.0in}}
\caption{From top to bottom, the probability distribution for
individual IMF multiplicities, the power-law exponent that describes
the IMF charge distributions, the breakup density, excess expansion
energy and time scale as a function of excitation energy for the
reaction of 8.0 GeV/c pions with gold nuclei.}
\label{Fig43}
\end{figure}

Figure \ref{Fig43}
summarizes several important features of the data that demonstrate a
distinct change in reaction mechanism at an excitation energy of E*/A
$\cong$ 4-5 MeV -- all of which support a phase-transition interpretation.
In the E*/A $\cong$ 4-5 MeV interval the emission of two or more IMFs
(accompanied by multiple thermal LCPs) becomes the dominant
disintegration mode.  From a power-law fit to the charge
distributions, it is found that the fraction of large clusters in an
average event is a maximum just above the transition excitation
energy, as predicted by theory.  Evidence for the conversion of extra
thermal energy into enhanced IMF kinetic energy appears near E*/A
$\sim$ 5 MeV, although the effect is small relative to heavy-ion
studies, where compressional heating may contribute.

Two unique results derived from ISiS are the evolution of the breakup
density and the disintegration time scale as a function of excitation
energy. The breakup density derived from analyses of the IMF
kinetic-energy spectra, provides evidence for emission from an
expanded/dilute source.  The derived breakup densities evolve from
normal density at low E*/A to a nearly constant value of
$\rho$/$\rho_0 \sim$ 0.3 near E*/A = 5 MeV and above, again consistent
with the theoretical predictions based on phase transition
assumptions.  The time scale for events with M$_{IMF}$ $>$ 2 
evolves from values characteristic of sequential statistical decay at
low excitation energy to times of the order of 20-50 fm/c for events
at E*/A $\sim$ 4 MeV and above, indicative of a near-simultaneous
decomposition mechanism.  

Similar to previous reports, the data also show evidence for a slope
change in the dependence of temperature on heat content, suggestive of
caloric curve behavior, although a distinct plateau is not observed.
Using a density-dependent Fermi-gas model to derive temperatures,
instead of the conventional isotope-ratio thermometer, a caloric curve
is obtained that is in good agreement with other results.  By gating
on bins of the preequilibrium spectra, it has also been possible to
derive a cooling curve for these hot systems.

\begin{figure}
\vspace{25mm}
\centerline{\psfig{file=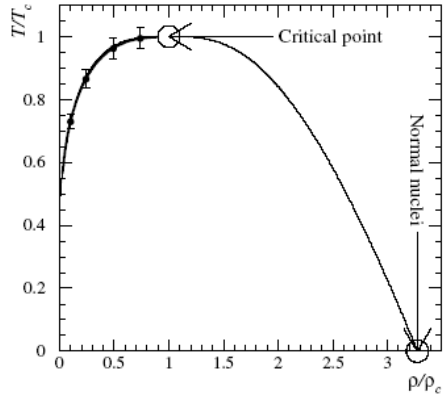,width=2.5in}}
\caption{The reduced density-temperature phase diagram: the thick line
is the calculated low density branch of the coexistence curve, the
points are selected calculated errors, and the thin lines are a fit to
the reflection of Guggenheim's equation.}
\label{Fig44}
\end{figure}

Finally, scaling-law fits to the IMF yield distributions provide
important confirmation of the statistical nature of the thermal events
observed in these studies.  A Fisher model analysis reveals excellent
scaling behavior up to E*/A $\sim$ 4 MeV, where sequential evaporative
emission dominates.  This result has been used to define the two-phase coexistence line for nuclear matter and serves as the basis
for derivation of the nuclear phase diagram shown in Fig. \ref{Fig44}.
A percolation model analysis that includes IMF secondary decay
corrections also describes the data well and indicates that at higher
excitaton energies, the data are described by a continuous phase
transition.  

In summary, all of the experimental evidence obtained in these studies
is consistent with the concept of a nuclear liquid-gas phase
transition.  However, the caveat must be added that given the finite
system size and very short time scales involved, the classical picture
of a phase transition must be invoked with caution.  It is perhaps
better to say that if a nuclear phase transition exists
in nuclei of a few hundred particles, this is what it looks like.

\section{Acknowledgments}

Over the fifteen year lifetime of the ISiS project numerous
individuals have provided invaluable support.  We wish to express our
sincere gratitude to the following for their help: computer support:
R.N. Yoder; ISiS design and construction: Andy Alexander, Kenny
Bastin, John Dorsett, Jack Ottarson, John Poehlman, Larry Sexton and
Lai Wan Woo; theoretical support: Alexandre Botvina, Wolfgang Bauer,
Pawel Danielewicz, C.B. Das, Jim Elliott, Bill Friedman, Subal das
Gupta, M. Kleine Berkenbusch, L.G. Moretto, and Viktor Toneev; and
undergraduates: Chris Powell and Greg Porter.  We also thank the
following for their advice and counsel: Joe Natowitz, Brian Serot and
Wolfgang Trautmann. We also thank John Vanderwerp, Kevin Komisarcik,
Steve Gushue, Lou Remsberg and Birger Back for their help in various
aspects of this project.  Diana McGovern played an indispensible role
in coordinating many aspects of the ISiS program and in manuscript
preparation.  Finally, we express our appreciation to accelerator
staffs at the Indiana University Cyclotron Facility, LNS Saclay and
Brookhaven AGS.

\end{document}